\documentclass{aa}

\usepackage{natbib,twoopt}
\usepackage[breaklinks=true]{hyperref}

\usepackage[varg]{txfonts}
\usepackage{graphicx}

\bibpunct{(}{)}{;}{a}{}{,}
\hypersetup{
    colorlinks,
    linkcolor=magenta,
    citecolor=blue,
    urlcolor=red}

\begin{document}

\title{Asymptotic power spectra and visibilities of damped mixed modes}
\subtitle{}

\author{Jonas M\"uller\inst{1,2}
\and Quentin Copp\'ee\inst{1}
\and Jordan Van Beeck\inst{1}
\and Tobias van Lier\inst{1,2}
\and Saskia Hekker\inst{1,2}}

\institute{Heidelberger Institut für Theoretische Studien, Schloss-Wolfsbrunnenweg 35, 69118 Heidelberg, Germany
\and Zentrum für Astronomie (ZAH/LSW), Heidelberg University, Königstuhl 12, 69117 Heidelberg, Germany}

%\titlerunning{}
\authorrunning{J. M\"uller et al.}

\date{Received <date> /Accepted <date>}

\abstract{Recent observational studies of red giant stars have estimated the visibility of their mixed oscillation modes, which is a proxy of the average energy of these modes.
Among other things, they demonstrated that although the damping rate of the oscillations in the core of many red giants appears to be negligible, other red giants exhibit high core damping rates that are sometimes consistent with the infinite value limit.
Up until now, it has not been possible to link the mixed mode visibilities to core damping rates in a quantitative way.
In this study, we use the progressive wave picture to derive an analytical function expressing the approximate resonance pattern of red giants up to a proportionality factor.
This function can model the influence of the damping on the oscillations, as well as take into account other effects such as mode asymmetries. 
In particular, this expression can be used to obtain a quantitative estimate for the visibility of mixed modes and to predict the detectability of mixed mode and multiplet signatures under different core damping rates.
Here, we conduct a parameter study to investigate how the damping processes affect these aspects.
We find that the visibility approaches the value expected for an infinite core damping rate already at finite values. Furthermore, we find that both the mixed mode and the multiplet signatures disappear at finite core damping rates. This implies that the observational characteristics of red giants with finite core damping rates can appear as if their core damping rate were infinite, providing an explanation for the observed populations. Moreover, we have used our method to quantitatively estimate the core damping rates of red giants with unusually low mixed mode amplitudes from their observed visibilities.
The analytical function describing the power spectrum developed in the present work is a flexible tool with many possible applications, which will be further explored in future studies.}

\keywords{asteroseismology $-$ stars: oscillations $-$ stars: interiors $-$ stars: evolution $-$ methods: analytical}

\maketitle

%\section{Title}\subsection{Title}\subsubsection{Title}\paragraph{Title}

\section{Introduction}

Solar-like oscillators are stars with a convective outer layer whose eigenmodes are intrinsically damped. Nevertheless, the turbulent motion of the convection near the star's surface can excite waves to observable amplitudes. On the main sequence, the oscillation modes detectable in solar-like oscillators are pressure modes. As the stars evolve into red giants, the density difference between the envelope and the core increases and eventually becomes so large that the multipolar oscillations take on a mixed character. This means that they behave like a pressure mode ("p-mode") in the outer layers and like a gravity mode ("g-mode") in the core of the star \citep[for a review, see][]{hekker+17}. The regions in which the p- and g-modes propagate are called "cavities" and are separated by a thin layer in which the oscillations become evanescent, known as the "evanescent zone". The mixed character of the mode is a consequence of part of the mode energy tunneling through the evanescent zone, enabling interaction between the two regions.
Since mixed modes are sensitive to the internal structure of the star, they have been successfully exploited to infer information about, for example, the core rotation rate \citep[e.g.,][]{beck+2012, mosser+12b, deheuvels+12, gehan+18, deheuvels+20,li+24}, the internal magnetic field \citep[e.g.,][]{li+22, li+23, deheuvels+23, hatt+24}, as well as the presence or absence of a convective core \citep[e.g.,][]{bedding+11, mosser+11,vrard+16}.

Mixed modes are usually described in an asymptotic framework, in which it is assumed that the wavelength of the oscillations is much smaller than the typical length scale of the variation in the stellar structure \citep{shibahashi79, tassoul80, unno+89}. Within this framework, a resonance condition can be constructed by matching the solutions for standing waves in the p-mode cavity with those in the g-mode cavity.
\citet{takata16b} shows that the resonance condition can also be obtained by considering the oscillations as progressive waves that are reflected and transmitted at the boundaries of the oscillation cavities. This approach is powerful because the energy loss caused by damping processes in the star can be described by partial reflection at the innermost and outermost boundaries. \citet{takata16b} uses this picture to derive an updated version of the resonance condition for mixed modes that takes into account the damping of the oscillations. 
The progressive wave picture is also adopted by \citet{pincon+takata22}, who generalize it to an arbitrary number of resonant cavities under the assumption of fully reflective boundary conditions (i.e., no damping). Their framework can be used to describe scenarios in which, for example, the inner g-mode cavity is divided into two parts by an intermediate convection zone. Furthermore, it allows the treatment of high-amplitude glitches that essentially divide a given cavity into two parts \citep{pincon19}.

There are two main types of damping processes that are expected to occur in every red giant star. The first is caused by the interaction of oscillations with convection in the upper region of the p-mode cavity, where the oscillation periods are comparable to the timescales of the most energetic convective eddies, so that the oscillations exert a feedback effect on the turbulence \citep[e.g.,][]{dupret+09,grosjean+14}. The second is radiative damping in the g-mode cavity, which is due to heat loss caused by fluctuations in the temperature gradient generated by the oscillations in the radiative core \citep[e.g.,][]{dziembowski77,vanHoolst+98,dziembowski+01}.
The radial modes are pure p-modes and are therefore only affected by the damping process in the p-mode cavity. Mixed modes are affected by the damping processes in both cavities, and their sensitivity to the individual processes depends on their mode inertia.
While the damping rate corresponding to interaction with convection is expected to remain of the same order of magnitude during the evolution of red giant stars \citep[e.g.,][]{vrard+18}, radiative damping should be negligible for the observable modes on the early red-giant branch, and increases strongly as the star evolves \citep[e.g.,][]{hekker+17}. However, observations of red giant stars have revealed a subgroup of stars with normal radial mode amplitudes but such low multipole mode amplitudes that the efficiency of the damping processes in their cores must be orders of magnitude higher than expected for radiative damping \citep[e.g.,][]{mosser+12,stello+16a,stello+16b,garcia+14, coppee+24}. 
In this context, \citet{fuller+15} suggest that such substantial energy losses could be caused by the interaction of the oscillations with a strong internal magnetic field, although the exact nature of this process is still the subject of intensive research \citep[e.g.,][]{lecoanet+17,loi20a, rui+fuller23,muller+25,david+25}.
Apart from these damping processes, which are all linear in nature, \citet{weinberg+19} and \citet{weinberg+21} investigated the impact of nonlinear interactions of the oscillation modes in RGB stars. \citet{weinberg+21} find that gravity-dominated mixed modes can be very susceptible to nonlinear damping on the late RGB. However, nonlinear damping does not seem to be able to explain the occurrence of stars with low multipole mode amplitudes in the early RGB, suggesting that it is not the dominant reason for the occurrence of these stars. Here, we focus on linear damping processes, and neglect nonlinear effects.

Adequate modeling of the feedback of these damping processes onto the oscillations is important for understanding how the energy loss affects the observable properties of the stellar oscillations.
One example of a quantity affected by these damping processes is the multipole mode visibility, which is a proxy of the ratio between the average energy of the multipole modes of a given spherical degree and the radial modes \citep{mosser+12}.
Another example is the presence or apparent absence of modes of mixed character in the power spectral density (PSD) of stars with unusually low multipole mode visibility.
On one hand, there are stars with low multipole mode visibilities for which no mixed modes have been detected within the observational uncertainties \citep[e.g.,][]{stello+16a, stello+16b}. On the other hand, there are stars with low multipole mode visibilities that show clear signs of mixed modes \citep{mosser+17, arentoft+17}. This raises the question of what properties the damping processes in the g-mode cavity must have so that the mixed modes are resolved and detectable in observations.
Frameworks for estimating the visibility and the height of damped mixed modes are currently only available within the limits of zero or infinitely large damping rates in the g-mode cavity \citep[e.g.,][]{shibahashi79,takata16b,mosser+17}. So far, it has therefore not been possible to link these observations with non-zero finite estimates for the damping rate in the g-mode cavity.

In this work, we used the progressive wave picture discussed by \citet{takata16b} and \citet{pincon+takata22} to derive an analytic expression for the theoretical resonance pattern of solar-like oscillators and the amplitudes of their oscillation modes up to a factor of proportionality. 
For this purpose, we have introduced an incident wave that drives the oscillations. In particular, we used the analogies of the waves in solar-like oscillators with light waves propagating in a Fabry-Pérot interferometer to derive an expression for the intrinsic resonances \citep[e.g.,][]{ismail+16}, which is proportional to the observable PSD of a star.
By integrating this expression over the oscillation frequency, we were able to calculate theoretical visibilities for damping processes of arbitrary efficiency, which we investigated over a large parameter space. In addition, we investigated how a damping processes in the g-mode cavity affects the detectability of mixed mode signatures in the PSD of red giant stars.
We also studied the detectability of mixed mode multiplets that can, for example, be caused by the rotation of the stellar core \citep[e.g.,][]{beck+2012}.
Finally, we used our formalism to provide quantitative estimates for the mode damping rates due to the damping processes occurring in the p- and g-mode cavities of the red giants in the sample of \citet{mosser+17}. These stars have dipole modes with low visibility but a distinct mixed character, which allowed us to relate their observed visibility to estimates of their damping rates in the g-mode cavity using our formalism.

This article is structured as follows. In Sect. \ref{sect: Progressive wave picture}, we introduce the progressive wave picture, in which an incident wave drives the oscillations. In Sect. \ref{sect: One and two cavity systems}, we discuss configurations with one and two cavities. In Sect. \ref{sect: Synthetic power spectrum}, we derive the analytical function for the theoretical power spectrum of the star and the visibility of multiple modes. In Sect. \ref{sect: Results}, we present our results. Finally, we conclude by summarizing and briefly discussing our findings in Sect. \ref{sect: Summary}.

\section{Progressive wave picture} \label{sect: Progressive wave picture}

In this study, we build on the theoretical framework developed by \citet{takata16b} and \citet{pincon+takata22} to infer the observable PSD of a solar-like oscillator up to a factor of proportionality. We adopt a slightly modified version of the notation of \citet{pincon+takata22}, which we present in this section.

\subsection{Wave function}

As shown by \citet{shibahashi79}, the radial part of the wave equation can be expressed within the asymptotic (WKBJ) and Cowling  \citep{cowling41} approximations in the following form:
\begin{gather}
    \frac{{\rm d}^2\Psi}{{\rm d}r^2} + k_{r}^2\ \Psi = 0.
    \label{eq: radial wave equation}
\end{gather}
In this expression, $\Psi$ is the wave function and the squared local radial wave number is given by
\begin{gather}
    k_{r}^2 = \frac{\omega^2}{c_{\rm s}^2} \left( \frac{N^2}{\omega^2} - 1 \right) \left( \frac{S_\ell^2}{\omega^2} - 1 \right),
    \label{eq: squared local radial wave number}
\end{gather}
where $\omega$ is the angular frequency, $N$ is the Brunt–Väisälä frequency, $S_\ell$ is the Lamb frequency corresponding to oscillations of the spherical degree $\ell$, and $c_{\rm s}$ is the adiabatic sound speed.
The waves resonate in the regions of the star where $k_{r}^2 > 0$, while they become evanescent if $k_{r}^2 < 0$. Inspection of Eq. \eqref{eq: squared local radial wave number} reveals that there are two kinds of asymptotic propagative solutions: p-modes (where $\omega^2 > N^2, S_\ell^2$) and g-modes (where $\omega^2 < N^2, S_\ell^2$). The regions of the star in which the waves propagate are the resonant cavities.

We assume that the solution of the radial part of the wave equation is a planar wave, which is given by
\begin{gather}
    \psi(r;\Bar{r}) = e^{\mathrm{i}\phi(r;\Bar{r})},
\end{gather}
where $\mathrm{i}$ is the imaginary unit, $r$ is the radial coordinate and $\Bar{r}$ is a reference radius in the corresponding cavity at which the phase is measured\footnote{In practice, we always place $\Bar{r}$ at the upper or lower boundary of the cavity under consideration.}.
The quantity $\phi$ is a complex phase, which we assume changes on a length scale comparable to $k_{r}^{-1}$.
We can then write the asymptotic solution of the radial part of the wave equation in a cavity as a superposition of planar waves:
\begin{gather}
    \Psi(r,t) \approx\left(a_{\rm pro}\ \psi(r;\Bar{r})\ + a_{\rm reg}\ \psi^\star(r;\Bar{r})\right) e^{-\mathrm{i}\omega t},
\end{gather}
where the $\star$-symbol indicates complex conjugation and $a_{\rm pro}$ ($a_{\rm reg}$) is the complex amplitude of the progressive (regressive) wave. Time is denoted by $t$.
The wave functions $\psi$ and $\psi^\star$ are linearly independent and constitute a basis for the solutions of the radial part of the wave equation \citep[e.g.,][]{pincon+takata22}.
Using Eq. \eqref{eq: radial wave equation}, the phase $\phi$ can be expanded and substituted, which yields \citep[e.g.,][]{gough07}
\begin{gather}
    \psi(r;\Bar{r}) \approx \frac{1}{\sqrt{k_{r}}} e^{\mathrm{i} \varphi(r;\Bar{r})}
\end{gather}
with
\begin{gather}
    \varphi(r; \Bar{r}) = \int_{\Bar{r}}^rk_{r}\ {\rm d}r^\prime.
\end{gather}
In the last equation, the independent variable is the upper limit of the integral.

Since progressive and regressive waves propagate in different directions, $\Psi$ can also be expressed as
\begin{gather}
    \Psi(r,t) = \left(a^\leftarrow _\pm\psi^\leftarrow_\pm(r) + a^\rightarrow _\pm\psi^\rightarrow_\pm(r)\right) e^{-\mathrm{i}\omega t},
    \label{eq: wave function arrows}
\end{gather}
where $a$ denotes the amplitude and the arrows indicate the direction of propagation (i.e., of the group velocity). 
In this expression, the subscript $_+$ ($_-$) refers to the fact that the reference radius $\Bar{r}$ is set equal to the upper (lower) boundary of the cavity under consideration. Therefore, either $_+$ or $_-$ must be selected and used consistently for all parameters.
The definition of $\psi^\rightarrow_\pm$ and $\psi^\leftarrow_\pm$ differs for p- and g-modes, since the phase and group velocities of pressure (gravity) waves have the same (opposite) direction.
For p-modes, the wave functions read
\begin{gather}
    \psi_{{\rm p},\pm}^\rightarrow(r) \equiv \psi(r;r_{{\rm p},\pm}), \quad \psi_{{\rm p},\pm}^\leftarrow(r) \equiv \psi^\star(r;r_{{\rm p},\pm}).
\end{gather}
Here, $r_{{\rm p},+}$ ($r_{{\rm p},-}$) denotes the radius of the upper (lower) boundary of the p-mode cavity.
The wave function for g-mode oscillations are given by
\begin{gather}
    \psi_{{\rm g},\pm}^\rightarrow(r) \equiv \psi^\star(r;r_{{\rm g},\pm}), \quad \psi_{{\rm g},\pm}^\leftarrow(r) \equiv \psi(r;r_{{\rm g},\pm}),
\end{gather}
where, $r_{{\rm g},+}$ ($r_{{\rm g},-}$) denotes the radius of the upper (lower) boundary of the g-mode cavity.

\subsection{Reflection and transmission}

The boundaries of the cavities are characterized by a reflection coefficient $R$ and a transmission coefficient $T$. In general, one has to distinguish the cases where the boundary is being approached from below (subscript $_+$) and from above (subscript $_-$).
However, using basic wave principles of time-reversal symmetry, linear superposition, and energy conservation, \citet{takata16b} shows that $|R| =|R_{\pm}|$ and $|T| =|T_{\pm}|$. Thus, we will neglect the subscript when referring to the modulus of a reflection or transmission coefficient.
In particular, the energy conservation implies that \citep{takata16b}
\begin{gather}
    |R|^2 + |T|^2 = 1
\end{gather}
must be fulfilled for each boundary.

The reflection of a wave at a boundary introduces a phase shift $\delta$.
If the wave approaches the boundary from below, the reflection coefficient at the upper boundary of the cavity can be expressed as \citep{takata16b}
\begin{gather}
    R_{+} = \frac{a_{+}^\leftarrow}{a_{+}^\rightarrow} = |R|\ e^{\mathrm{i}\delta} = e^{\mathrm{i}(\delta + \mathrm{i}2\mu)}.
\end{gather}
Here, we have introduced the amplitude modification factor
\begin{gather}
    \mu = -\frac{1}{2}\ln(|R|),
    \label{eq: amplitude modification factor}
\end{gather}
which vanishes in the fully reflective case (i.e., if $|R| = 1$)\footnote{\citet{takata16b} defines $\mu$ as a complex quantity that depends on $R$ rather than $|R|$. In the definition used in this work, the complex phase of $R$ is contained in $\delta$, making $\mu$ a real parameter.}.
If the wave approaches the boundary from above, the reflection coefficient at the lower boundary of the cavity can be expressed as \citep{takata16b}
\begin{gather}
    R_{-} = \frac{a_{-}^\rightarrow}{a_{-}^\leftarrow} = |R|\ e^{\mathrm{i}(\pi-\delta)} = e^{\mathrm{i}(\pi-\delta + \mathrm{i}2\mu)}
\end{gather}
For simplicity, we assume that there is no phase lag induced by the transmission of the wave.

\subsection{Amplitude conversion factors}

The wave function must be independent of the choice of the reference radius of the phase $\Bar{r}$.
This allows us to obtain amplitude conversion factors that measure the phase shift that the wave acquires as it traverses the cavity once by comparing $\Psi(r,t)$ with $\Bar{r}=r_-$ with a function that describes the same wave with $\Bar{r}=r_+$ in the same cavity. The conversion factors read
\citep{pincon+takata22}:
\begin{gather}
    a_{-}^\rightarrow = e^{-\mathrm{i} \Theta}\ a_{+}^\rightarrow, \\
    a_{-}^\leftarrow = e^{\mathrm{i} \Theta}\ a_{+}^\leftarrow
\end{gather}
where
\begin{gather}
    \Theta = \pm\int_{r_-}^{r_+} k_{r}\ {\rm d}r.
    \label{eq: theta}
\end{gather}
In this equation, $r_+$ ($r_-$) is the upper (lower) boundary of the corresponding cavity. The plus (minus) sign in front of the integral must be used for p-modes (g-modes) due to the differing orientation of the group velocity.

Equation \eqref{eq: theta} can be calculated explicitly for p- and g-modes using Eq. \eqref{eq: squared local radial wave number}.
For p-modes, we make the common assumption $\omega^2 \gg N^2, S_\ell^2$. Therefore, we obtain
\begin{gather}
    \Theta_{\rm p} \approx \frac{\pi\ \nu}{\Delta\nu},
    \label{eq: theta_p}
\end{gather}
where we identified the integral of the inverse adiabatic sound speed over the radius with the large frequency separation $\Delta\nu$ \citep[e.g., equation (7) of][]{hekker+17}\footnote{The large frequency separation $\Delta\nu$ is technically defined as the integral of the inverse adiabatic sound speed from the center of the star ($r=0$) to its surface ($r=R_*$), which roughly corresponds to the integral over the p-mode cavity.}.
For the g-modes, we assume $\omega^2 \ll N^2, S_\ell^2$, which leads to
\begin{gather}
    \Theta_{\rm g} \approx -\frac{\pi}{\Delta\Pi_\ell \ \nu}.
\end{gather}
In this expression, $\Delta\Pi_\ell$ denotes the asymptotic period spacing of the modes of spherical degree $\ell$ \citep[e.g., equations (11) and (12) of][]{hekker+17}.

\section{Systems with one and two cavities} \label{sect: One and two cavity systems}

\subsection{One cavity} \label{sect: one cavity system}

\begin{figure}[]
    
    \resizebox{\hsize}{!}{\includegraphics{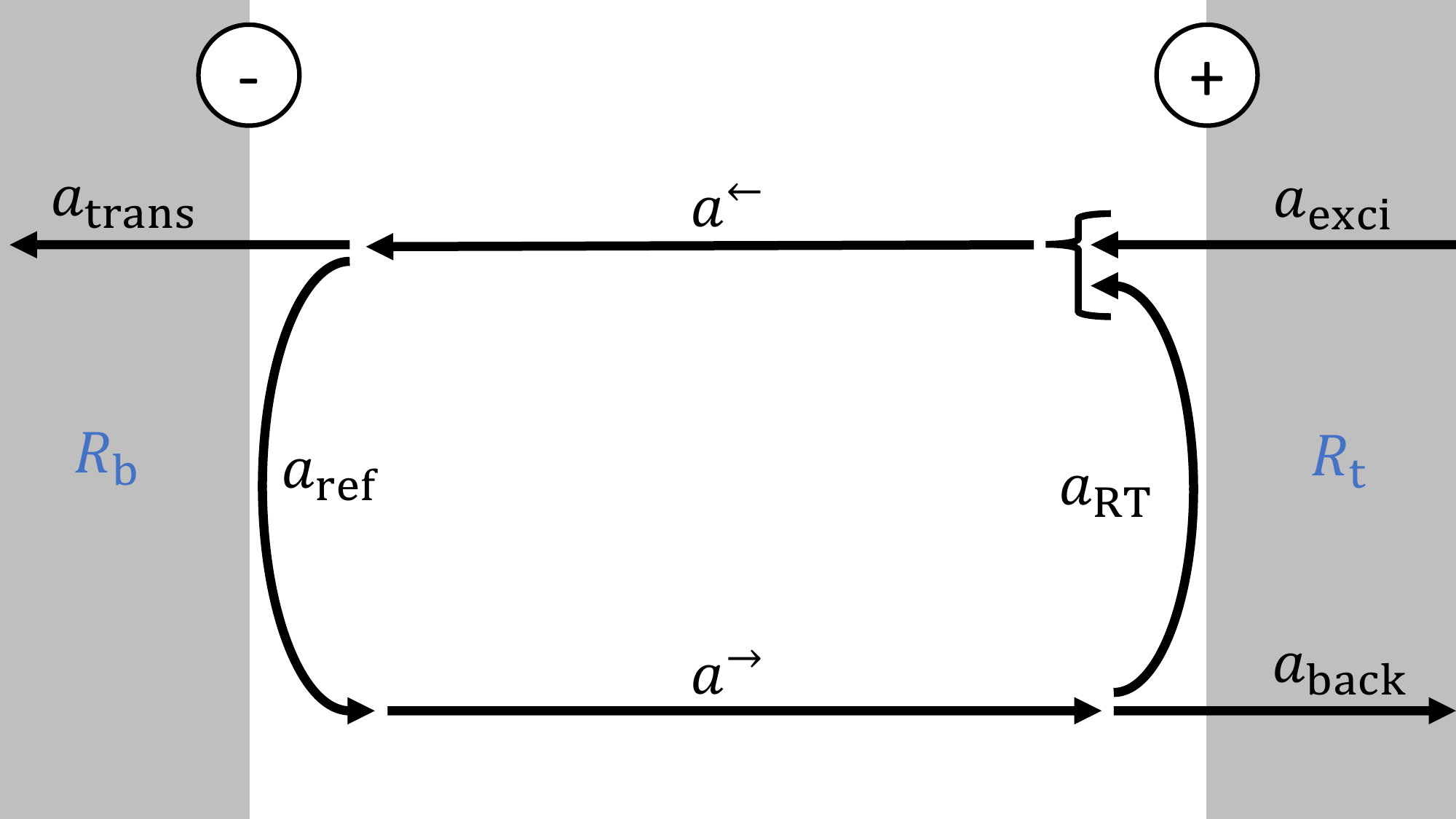}}
    \caption{Sketch of the single-cavity system. The curly bracket indicates interference.}
    \label{fig: sketch 1cavity}
\end{figure}

As a first step, we consider a system consisting of a single cavity confined by an upper boundary at the top (subscript $_{\rm t}$) and a lower boundary at the bottom of the cavity (subscript $_{\rm b}$). The setup is shown in Fig. \ref{fig: sketch 1cavity}. The oscillations are excited by a wave with the amplitude $a_{\rm exci}$ coming from an external source, which enters the system through the upper boundary. The oscillations are damped by the parts of the waves that are transmitted through the upper and lower boundaries and thus escape from the system. These correspond to the amplitudes $a_{\rm back}$ and $a_{\rm trans}$. Furthermore, we assume that the system has reached a steady state.

\subsubsection{Resonance condition}

The amplitude of the downward-traveling wave at the upper boundary $a^\leftarrow_+$ and the amplitude of the excitation $a_{\rm exci}$ are related to each other by the following expression:
\begin{gather}
    a^\leftarrow_+ = a_{\rm exci} + a_{\rm RT} = a_{\rm exci} + R_{{\rm b},-}\  R_{{\rm t},+}\ e^{\mathrm{i}2\Theta}\ a^\leftarrow_+.
\end{gather}
The subscript $_{\rm RT}$ corresponds to the amplitude of the part of the wave that has completed a round trip in the cavity, which is used to substitute it on the right side of the above equation.
The above expression can be used to derive the internal resonance enhancement factor $A_1$, which constitutes an Airy distribution and measures the response of the system to the excitation \citep{ismail+16}. It is given by
\begin{gather}
    A_1 = \frac{|a^\leftarrow_+|^2}{|a_{\rm exci}|^2} = \left|1 - e^{\mathrm{i}2\left(\Theta + \frac{\pi}{2}-\frac{\delta_{\rm b}}{2}+\frac{\delta_{\rm t}}{2} + \mathrm{i}(\mu_{\rm b} + \mu_{\rm t})\right)}\right|^{-2}.
    \label{eq: IREF 1 cavity}
\end{gather}

From a theoretical point of view, resonant oscillation modes are strictly speaking idealized standing waves that only require an initial kick that took place in an infinitely distant past to oscillate. This means that they do not need a continuous energy supply, which corresponds to $a_{\rm exci} = 0$, causing the internal resonance enhancement factor $A_1$ to diverge at the eigenfrequencies of the resonant modes.
Enforcing this, we obtain two separable resonance conditions for the real and imaginary parts of $\Theta$:
\begin{gather}
    \Theta_\Re - \pi\epsilon = n \pi, \label{eq: resonance condition 1 cavity real}\\
    \Theta_\Im + \mu_{\rm b} + \mu_{\rm t} = 0.
    \label{eq: resonance condition 1 cavity imag}
\end{gather}
Here, $n$ is an integer, and the phase shift $\epsilon$ is given by
\begin{gather}
    \epsilon \equiv -\frac{1}{2} +\frac{\delta_{\rm b}}{2\pi}- \frac{\delta_{\rm t}}{2\pi}.
\end{gather}
The subscript $_\Re$ ($_\Im$) indicates the real (imaginary) part throughout this paper.
According to Eqs. \eqref{eq: resonance condition 1 cavity real} and \eqref{eq: resonance condition 1 cavity imag}, the eigenfrequencies of the oscillation modes become complex when $|R_{\rm t}| < 1$, $|R_{\rm b}| < 1$, or both, thereby compensating for the loss of energy. 

Although treating stellar oscillations as modes is a useful mathematical framework, idealized modes do not require a source of excitation. In solar-like oscillators however, the energy loss of the waves is compensated for by stochastic excitation in the convective envelope, which provides a driving mechanism for the oscillations. Therefore, the oscillations in the star can also be understood as a wave with a real frequency that is excited and damped by incident and outgoing waves of the same real frequency, rather than considering idealized modes with complex eigenfrequencies. We discuss this interpretation further in Sect. \ref{sect: Synthetic power spectrum}.

\subsubsection{Physical meaning of the amplitude modification factor}

The amount of damping of a wave is commonly quantified using a damping rate $\eta$. The time dependence of the wave then becomes $e^{-\mathrm{i}\omega t - \eta t}$, where $\eta$ is a real number. This additional time-dependent factor influences the modulus of the amplitude as the wave propagates (i.e., |$\Tilde{a}| = |a| |e^{-\eta \cdot t}|$). In that sense, $\eta$ has a similar effect on the amplitudes as the amplitude modification factor $\mu$. To compare these two parameters, we now assume $\eta\neq0$ and fully reflective boundary conditions (i.e., $\mu_{\rm b} = \mu_{\rm t} = 0$).

The time required for the wave to travel back and forth through the cavity (i.e., the round-trip time) is twice as long as the crossing time, which is defined as
\begin{gather}
    t_{\rm cross} = \pm\int_{r_-}^{r_+} \frac{{\rm d}r}{c},
    \label{eq: crossing time}
\end{gather}
Here, $c$ is the propagation speed of the wave packet (i.e., the group velocity) and the plus (minus) sign is appropriate for the p-modes (g-modes). Accordingly, we obtain the following amplitude moduli:
\begin{gather}
    |\Tilde{a}^\leftarrow_+ |= |a^\leftarrow_+| , \quad |\Tilde{a}_{\rm exci}| = |a_{\rm exci}|, \quad |\Tilde{a}_{\rm RT}| = \left|a_{\rm RT}\ e^{-\eta \cdot 2t_{\rm cross}}\right|.
\end{gather}
Using these amplitude moduli, the internal resonance enhancement factor can be written as
\begin{gather}
    \Tilde{A}_1 = \frac{|\Tilde{a}^\leftarrow_+|^2}{|\Tilde{a}_{\rm exci}|^2} = \left|1 - e^{\mathrm{i}2(\Theta - \pi\epsilon + \mathrm{i}\eta \cdot t_{\rm cross})}\right|^{-2},
\end{gather}
which is very similar to Eq. \eqref{eq: IREF 1 cavity}.

Comparison of $A_1$ in Eq. \eqref{eq: IREF 1 cavity} and $\Tilde{A}_1$ implies that the energy loss through the boundaries with no intrinsic damping rate can be described as the energy loss of a wave with a nonzero intrinsic damping rate under fully reflective boundary conditions if\footnote{The left side of Eq. \eqref{eq: amplitdue modification vs. eta} refers to the case with energy loss due to partial reflection at the boundaries, while the middle and right sides refer to fully reflective boundary conditions and a intrinsic nonzero damping rate.}
\begin{gather}
    \mu_{\rm b} + \mu_{\rm t}= \eta \cdot t_{\rm cross} \equiv (\eta_{\rm b} + \eta_{\rm t}) \cdot t_{\rm cross}.
    \label{eq: amplitdue modification vs. eta}
\end{gather}
Therefore, we identify the amplitude modification factor $\mu$ as the product of the damping rate of its corresponding damping process and the crossing time of the considered cavity.

\subsubsection{Pressure and gravity modes} \label{sect: pressure and gravity modes}

Now let us return to the theoretical picture shown in Fig. \ref{fig: sketch 1cavity} and apply it to the p- and g-modes, respectively.
First, we consider p-modes and assume that the energy loss occurs at the upper boundary. Therefore, we set $\mu_{\rm b} = 0$ and $\mu_{\rm t} = \mu_{\rm p}$. Using Eq. \eqref{eq: resonance condition 1 cavity imag}, the remaining amplitude modification factor can then be expressed as \citep[see also][]{takata16b}
\begin{gather}
    \mu_{\rm p} = -\frac{\pi\ \nu_\Im}{\Delta\nu} = \frac{\eta_{\rm p}}{2\ \Delta\nu} = \eta_{\rm p} \cdot t_{\rm p,cross},
    \label{eq: imaginary nu pressure}
\end{gather}
where $\eta_{\rm p}$ is the damping rate corresponding to the energy loss in the p-mode cavity and $t_{\rm p,cross}$ is the crossing time of the p-mode cavity.
If, on the other hand, we allow $\mu_{\rm b}$ to be greater than zero, the situation is equivalent to an arrangement with two cavities, in which any contribution that enters the second cavity is lost. This special scenario has been thoroughly discussed in the literature \citep[e.g.,][]{osaki77,dziembowski77,shibahashi79,fuller+15,takata16b}. In this case, $\mu_{\rm b}$ thus measures the amount of interaction between the p-mode cavity and the second cavity.
We summarize the interrelations between the parameters describing the efficiency of the damping processes and the underlying assumptions in Appendix \ref{app: Parameters describing mode damping}.

Following a similar argument for the g-modes, we now assume that the energy loss occurs at the lower boundary, as one would expect for the g-mode cavity in a red giant star \citep[e.g.,][]{takata16b}. Therefore, we set $\mu_{\rm b} = \mu_{\rm g}$ and $\mu_{\rm t} = 0$. This yields
\begin{gather}
    \mu_{\rm g} = -\frac{\pi\ \nu_\Im}{\Delta\Pi_\ell\ |\nu|^2} = \frac{\eta_{\rm g}}{2\ \Delta\Pi_\ell\ |\nu|^2} = \eta_{\rm g} \cdot t_{\rm g,cross}.
    \label{eq: imaginary nu gravity}
\end{gather}
Here, $\eta_{\rm g}$ is the damping rate corresponding to the energy loss in the g-mode cavity and $t_{\rm g,cross}$ is the crossing time of the g-mode cavity.

\subsection{Two cavities} \label{sect: two cavity system}

\begin{figure}[]
    
    \resizebox{\hsize}{!}{\includegraphics{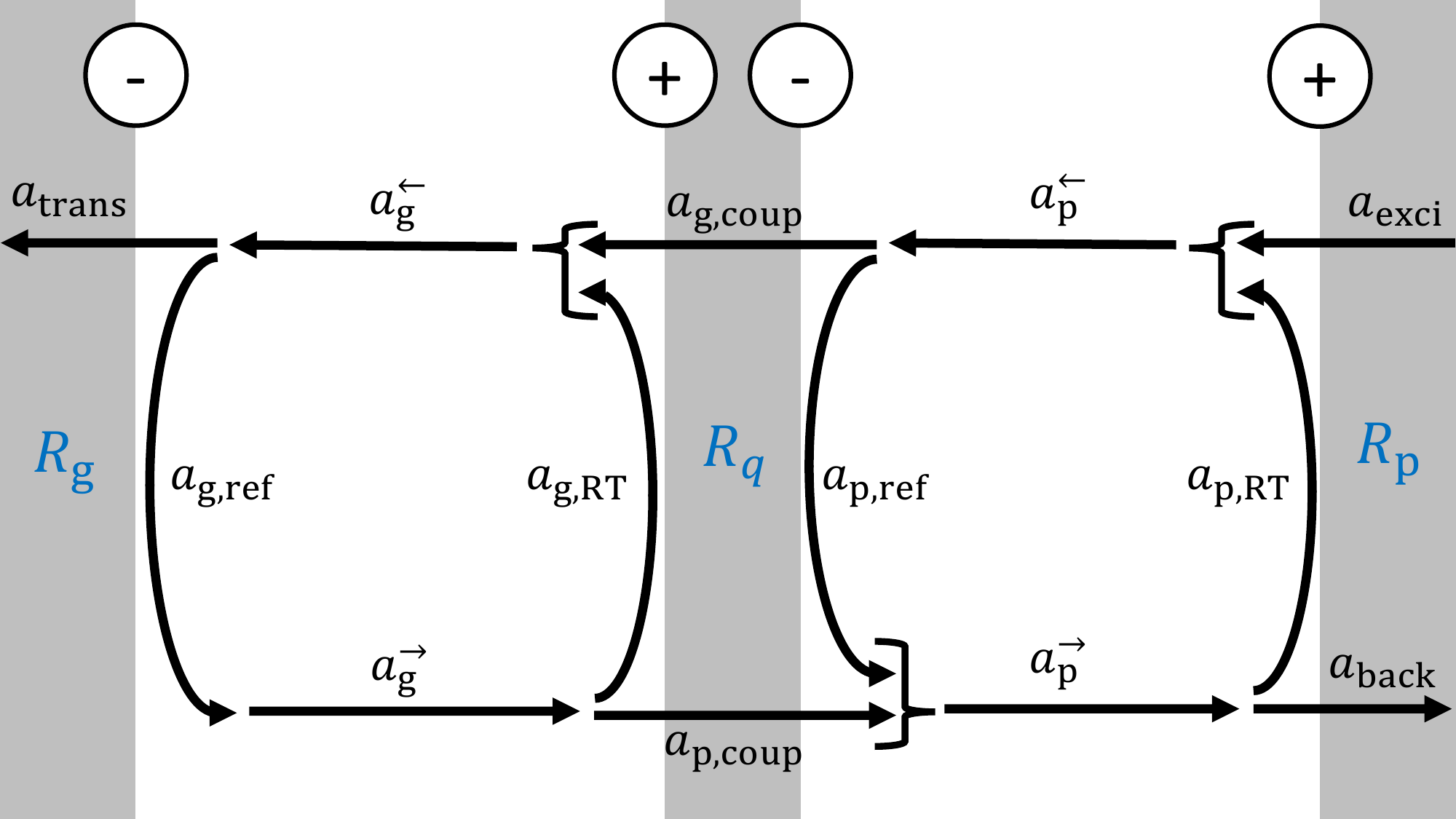}}
    \caption{Sketch of the two-cavity system. Curly brackets indicate interference.}
    \label{fig: sketch 2cavities}
\end{figure}

Next, we consider a system consisting of two cavities coupled by the leakage of energy through a shared intermediate boundary (the evanescent zone, or "EZ" for short; subscript $_q$). The arrangement is shown in Fig. \ref{fig: sketch 2cavities}. In particular, we assume that the upper cavity is a p-mode cavity (subscript $_{\rm p}$) and the lower cavity is a g-mode cavity (subscript $_{\rm g}$). Excitation occurs at the upper boundary of the p-mode cavity, and energy losses can occur both at the upper boundary of the p-mode cavity and at the lower boundary of the g-mode cavity. This system is reminiscent of the mixed mode oscillations that occur in red giant stars \citep[see also][]{takata16b}.

\subsubsection{Infinite reflections in the gravity mode cavity}

Similar to the single-cavity system, our goal is to find an expression for the internal resonance enhancement factor. Therefore, we must first derive an expression for the amplitude $a^\rightarrow_{{\rm p},-}$ as a function of $a^\leftarrow_{{\rm p},-}$, which now contains an additional contribution from the energy leaking from the g- into the p-mode cavity.
Taking into account the contributions caused by the infinite number of reflections in the g-mode cavity, the outward-propagating wave in the p-mode cavity can be written as
\begin{gather}
    a^\rightarrow_{{\rm p},-} = a_{\rm p,ref} + a_{\rm p,coup}  = \overbrace{R_{q,-}}^{\substack{\text{reflection at lower boundary} \\ \text{of p-mode cavity}}} a^\leftarrow_{{\rm p},-} \notag\\
    \quad + \underbrace{|T_q|}_{\substack{\text{transm.} \\ \text{through} \\ \text{EZ}}} \underbrace{R_{{\rm g},-}\ e^{\mathrm{i}2\Theta_{\rm g}}}_{\substack{\text{reflection at} \\ \text{lower boundary} \\ \text{of g-mode cavity}}}\underbrace{\sum_k^\infty \left(R_{q,+}\ R_{{\rm g},-}\ e^{\mathrm{i}2\Theta_{\rm g}}\right)^k}_{\substack{\text{infinite number of back and forth} \\ \text{reflections in g-mode cavity}}} \underbrace{|T_q|}_{\substack{\text{transm.} \\ \text{back} \\ \text{through} \\ \text{EZ}}} a^\leftarrow_{{\rm p},-}.
    \label{eq: upward vs downward 2 cavities}
\end{gather}
It can be shown that, after some manipulations, this equation can be expressed as a single exponential function \citep[for a more detailed derivation, see section 3.1. of][]{pincon+takata22}
\begin{gather}
    a^\rightarrow_{{\rm p},-} = e^{\mathrm{i}2\left(\frac{\pi}{2} - \frac{\delta_q}{2} +\Phi_{\rm g}\right)}\  a^\leftarrow_{{\rm p},-},
    \label{eq: ampl up down relation}
\end{gather}
where we defined
\begin{gather}
    \Phi_{\rm g} \equiv \arctan\left(q \cot\left( -\Theta_{\rm g} - \frac{\pi}{2} - \pi\epsilon_{\rm g} - \mathrm{i}\mu_{\rm g}\right)\right)
\end{gather}
and
\begin{gather}
    \epsilon_{\rm g} \equiv \frac{\delta_q}{2\pi} -\frac{\delta_{\rm g}}{2\pi}.
\end{gather}
In addition, we introduced the coupling strength \citep[see][]{shibahashi79, tassoul80, unno+89,takata16a}, which is given by \citep{takata16b}
\begin{gather}
    q = \frac{1 - |R_q|}{1 + |R_q|}.
    \label{eq: coupling strength}
\end{gather}
The coupling strength is a proxy for the modulus of the reflection coefficient of a boundary, just like the amplitude modification factor $\mu$.

\subsubsection{Resonance condition}

Using Eq. \eqref{eq: ampl up down relation}, we can express the amplitude of the downward-traveling wave in the p-mode cavity as a function of the amplitude of the excitation $a_{\rm exci}$:
\begin{gather}
    a^\leftarrow_{{\rm p},+} = a_{\rm exci} + a_{\rm p,RT} = a_{\rm exci} +  R_{{\rm p},+}\ e^{\mathrm{i}2\Theta_{\rm p}}\ e^{\mathrm{i}2\left(\frac{\pi}{2} - \frac{\delta_q}{2} +\Phi_{\rm g}\right)}\ a^\leftarrow_{{\rm p},+}.
\end{gather}
Analogously to the arrangement with one cavity, we define the internal resonance enhancement factor of the two-cavity system as
\begin{gather}
    A_2 = \frac{|a^\leftarrow_{{\rm p},+}|^2}{|a_{\rm exci}|^2} = \left|1 - e^{\mathrm{i}2\left(\Theta_{\rm p}- \pi\epsilon_{\rm p}  + i\mu_{\rm p} + \Phi_{\rm g}\right)}\right|^{-2}.
    \label{eq: resonance enhancement factor 2}
\end{gather}
The phase shift in the p-mode cavity is defined as
\begin{gather}
    \epsilon_{\rm p} \equiv -\frac{1}{2} +\frac{\delta_{q}}{2\pi}- \frac{\delta_{\rm p}}{2\pi}.
\end{gather}
Comparison of $A_1$ and $A_2$ reveals that the amplitude modification factor of the lower boundary $\mu_{\rm b}$ in the single-cavity system has been replaced by the term $-\mathrm{i}\Phi_{\rm g}$, which accounts for the interaction with the g-mode cavity. Note that, contrary to $\mu_{\rm b}$, $\Phi_{\rm g}$ is generally complex.

Similar to the single-cavity case, the resonance condition can be obtained as the condition under which $A_2$ diverges, which yields
\begin{gather}
    \Theta_{\rm p} - \pi\epsilon_{\rm p}  + \mathrm{i}\mu_{\rm p} + \Phi_{\rm g} = n\pi,
    \label{eq: resonance condition 2 cavities with n}
\end{gather}
where $n$ is an integer. This expression can be rewritten in a more recognizable form:
\begin{gather}
    \cot\left( -\Theta_{\rm g} - \pi\epsilon_{\rm g} -\mathrm{i}\mu_{\rm g} \right) \tan\left( \Theta_{\rm p} - \pi\epsilon_{\rm p}+\mathrm{i}\mu_{\rm p} \right) = q.
    \label{eq: resonance condition 2 cavities}
\end{gather}
Equation \eqref{eq: resonance condition 2 cavities} is equivalent to the resonance condition derived by \citet{takata16b} in the case of leakage of mode energy through the outer boundaries.
The minus sign in front of $\Theta_{\rm g}$ is due to the orientation of the group velocity in the g-mode cavity (see Eq. \eqref{eq: theta}). The eigenfrequencies of the resonant modes are given by the roots of Eq. \eqref{eq: resonance condition 2 cavities}, which are complex except for totally reflective boundary conditions.

\section{Synthetic power spectrum} \label{sect: Synthetic power spectrum}

The PSD of a star is an important observational constraint. Intrinsically damped oscillation modes appear in the PSD as Lorentz functions to a good approximation, provided that the frequency resolution is good enough (i.e., much smaller that the linewidth of the modes). To compare theoretical predictions with observations, it can thus be insightful to generate synthetic PSDs for a preselected set of stellar parameters or for a numerical stellar model. Usually, this is done by finding the eigenfrequencies of the oscillation modes using the resonance condition and subsequently using these frequencies as inputs for the Lorentz functions. The total PSD is then calculated as the sum of the Lorentzians of all oscillation modes (see cyan dotted lines in Figs. \ref{fig: IRE_Lorentz_1cavity} and \ref{fig: IRE_Lorentz_2cavities}). 

While this approach works well for a single-cavity system and for a two-cavity system with totally reflective boundary conditions, it can be quite involved for an arrangement with two cavities and energy loss at the boundaries. On one hand, this is because it can be challenging to reliably find all the roots of a discontinuous function in a vertical band of the complex plane \citep[e.g.,][]{goldstein+townsend20}. On the other hand, there is currently no expression for determining the relative peak height of a mixed mode experiencing damping in the g-mode cavity, unless the corresponding damping rate $\eta_{\rm g}$ is zero or infinite.

\subsection{Normalized power spectrum}

\begin{figure}[]
    
    \resizebox{\hsize}{!}{\includegraphics{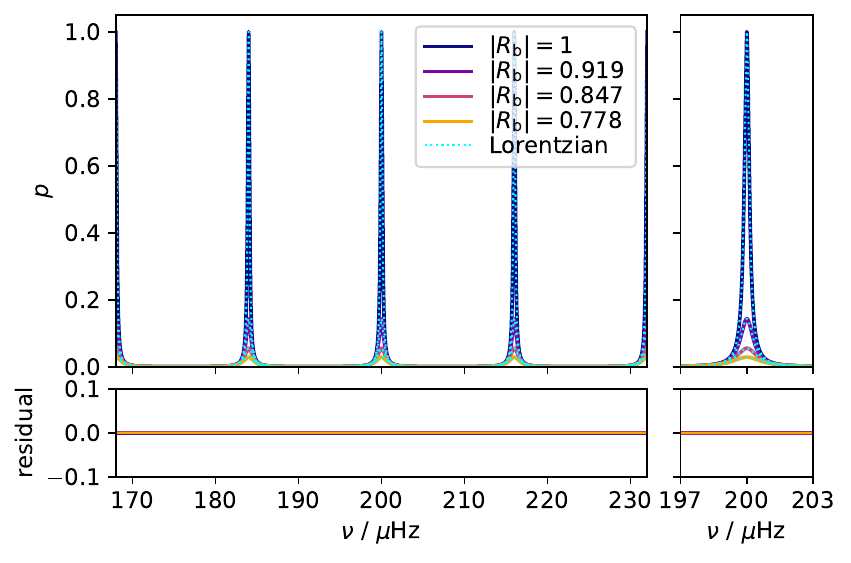}}
    \caption{Normalized power spectrum as a function of frequency for the set of stellar parameters A (see Table \ref{tab: sets of stellar parameters}) with $|R_{\rm t}|=0.95$ for different values of $|R_{\rm b}|$ in a single-cavity system. Colored solid lines show $p$ calculated using Eq. \eqref{eq: normalized power spectrum 1 cavity}. Cyan dotted lines show the normalized power spectrum a sum of Lorentzians whose positions and widths were calculated using the resonance condition. The frequency resolution is selected so that all peaks are resolved. In the bottom row, we show the residual of each $p$ and its corresponding sum of the Lorentzian functions.}
    \label{fig: IRE_Lorentz_1cavity}
\end{figure}
\begin{figure}[]
    
    \resizebox{\hsize}{!}{\includegraphics{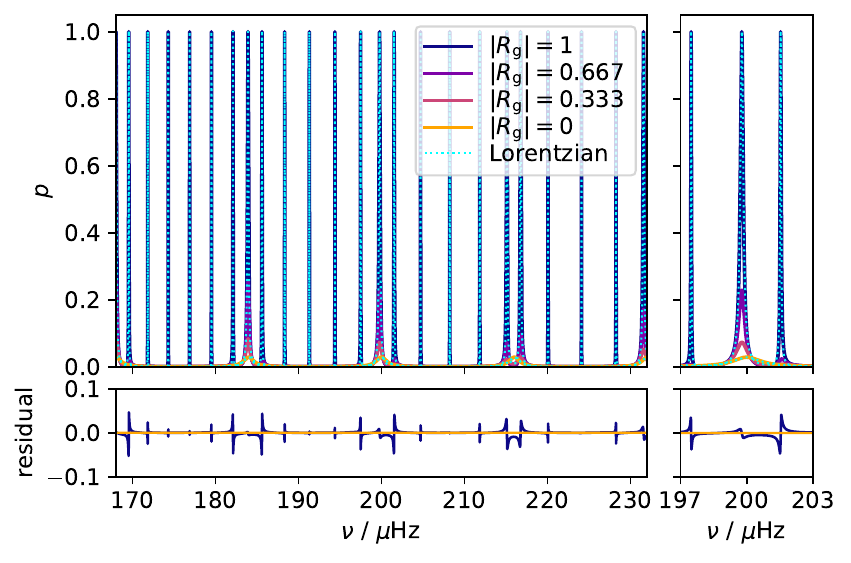}}
    \caption{Same as Fig. \ref{fig: IRE_Lorentz_1cavity} calculated with Eq. \eqref{eq: normalized power spectrum 2 cavities} for different values of $|R_{\rm g}|$ in a two-cavity system. Lorentzians are only shown for $|R_{\rm g}|=1$ (i.e., no damping in the g-mode cavity) and 0 (i.e., complete loss of energy in the g-mode cavity), since the peak height is unconstrained otherwise (see main text). The case $|R_{\rm g}|=0$ is identical to the case $|R_{\rm b}|=0.778$ in Fig. \ref{fig: IRE_Lorentz_1cavity}.}
    \label{fig: IRE_Lorentz_2cavities}
\end{figure}

To mitigate both of the aforementioned difficulties, we propose an alternative way of calculating synthetic PSDs. 
In this section, we compare the amplitudes of the oscillations in the p-mode cavity with that of the excitation, so that $a_{\rm exci}$ must be greater than zero (except when $|R_{\rm p}|=1$). This is necessary because the amplitudes in the p-mode cavity depend on $a_{\rm exci}$. In particular, this means that we do not consider idealized oscillation modes with complex eigenfrequencies as done in Sect. \ref{sect: Progressive wave picture}, but rather waves with real frequencies that are driven and damped by incident and outgoing waves of the same real frequency.
We thus assume that $\nu$ lies on the real axis. Because of that, we have omitted the subscripts $_+$ and $_-$ in the modulus of the amplitudes.
Nevertheless, we will continue to use the term “mode” because it is standard in the literature and the waves can formally be understood as modes with complex frequencies.

For waves with real frequencies, the power spectrum is approximately proportional to the sum of the squared modulus of the amplitude of the downward- and upward-traveling waves in the p-mode cavity scaled by a frequency-dependent factor $C_\omega$ (see Appendix \ref{app: Power spectral density} for a justification):
\begin{gather}
    P \propto C_\omega\left(|a_{\rm p}^{\leftarrow}|^2 + |a_{\rm p}^{\rightarrow}|^2\right).
\end{gather}
The exact values of the amplitudes $a_{\rm p}^{\leftarrow}$ and $a_{\rm p}^{\rightarrow}$ are functions of the amplitude of the excitation $a_{\rm exci}$, which is unknown, so that $P$ cannot be determined. Therefore, we normalize the power spectrum based on the signal that would be observable if only the downward-traveling wave corresponding to the excitation were present. In short, we normalize based on the power spectrum that would occur if there were no boundaries in the star, which can be expressed as
\begin{gather}
    P_{\rm exci} \propto C_\omega\ |a_{\rm exci}^{\leftarrow}|^2.
\end{gather}

We introduce the normalized power spectrum $p$ as
\begin{gather}
    p \equiv \frac{1}{\mathcal{H}_{\rm p}} \frac{P}{P_{\rm exci}} = \frac{1}{\mathcal{H}_{\rm p}} \frac{|a_{\rm p}^{\leftarrow}|^2 + |a_{\rm p}^{\rightarrow}|^2}{|a_{\rm exci}|^2}.
    \label{eq: normalized power spectrum}
\end{gather}
In this expression, $\mathcal{H}_{\rm p}$ is the height of the peaks that only experience energy loss at the upper boundary of the p-mode cavity (see Eq. \eqref{eq: peak height}).
Since we divide by $\mathcal{H}_{\rm p}$, the height of the peaks that only experience energy loss at the upper boundary of the p-mode cavity is equal to one. Waves that experience additional damping appear with a height of less than one. In other words, the frequency-dependence of the excitation is canceled out by the factor $1/P_{\rm exci}$, while the frequency-dependence of the damping in the p-mode cavity is eliminated by $1/\mathcal{H}_{\rm p}$.
Nevertheless, $p$ retains a frequency-dependence due to the differing interference patterns of waves with different $\nu$.
In this work, we limit ourselves to constant values of the reflection coefficients, such that $1/\mathcal{H}_{\rm p}$ becomes a normalization constant.
Since Eq. \eqref{eq: normalized power spectrum} describes the observable power spectrum up to a function that depends on the frequency dependence of the excitation and damping at the upper boundary of the p-mode cavity, it can be used to compare the relative heights of the peaks of waves that are subject to different amounts of damping at the lower boundary of the p-mode cavity in the single-cavity system or in the g-mode cavity in the two-cavity system.

In a single-cavity system, the normalized power spectrum can be written as a function of the internal resonance enhancement factor $A_1$:
\begin{gather}
   p = \left(1 + |R_{\rm b}|^2\right) \frac{A_1}{\mathcal{H}_{\rm t}}.
   \label{eq: normalized power spectrum 1 cavity}
\end{gather}
Here, $\mathcal{H}_{\rm t}$ denotes the peak height of the peaks of waves that do not lose energy at the lower boundary of the cavity, which is a function of $|R_{\rm t}|$ (see below).
In the two-cavity system described in Sect. \ref{sect: two cavity system}, the normalized power spectrum is given by
\begin{gather}
   p = \left(1 + e^{-4(\Phi_{\rm g})_\Im}\right) \frac{A_2}{\mathcal{H}_{\rm p}},
   \label{eq: normalized power spectrum 2 cavities}
\end{gather}
where $e^{-(\Phi_{\rm g})_\Im}$ takes the role of an effective reflection coefficient that accounts for the energy transmitted back from the g- to the p-mode cavity.

The peak height $\mathcal{H}_{\rm p}$ can be calculated by assuming constructive interference of the waves in the p-mode cavity and that the waves only experience energy losses at the uppermost boundary of the system. It is given by
\begin{gather}
    \mathcal{H}_{\rm p} = \frac{2}{\left(1 - |R_{\rm p}| \right)^2}.
    \label{eq: peak height}
\end{gather}
The peak height $\mathcal{H}_{\rm t}$, which is important for the arrangement with one cavity (see Eq. \eqref{eq: normalized power spectrum 1 cavity}), can also be estimated using Eq. \eqref{eq: peak height}, by replacing the subscript $_{\rm p}$ with $_{\rm t}$.

We verify this approach by comparing the normalized power spectrum to a sum of Lorentz functions in Figs. \ref{fig: IRE_Lorentz_1cavity} and \ref{fig: IRE_Lorentz_2cavities} for an arrangement with one and two cavities, respectively. In the system with one cavity (Fig. \ref{fig: IRE_Lorentz_1cavity}), the two approaches agree perfectly in all cases. In the system with two cavities, the two approaches only agree when $R_{\rm g}=0$. If $|R_{\rm g}|=1$, we observe in Fig. \ref{fig: IRE_Lorentz_2cavities} that both the position and the width of the Lorentzians agree very well with the profile calculated using Eq. \eqref{eq: normalized power spectrum 2 cavities}. 
However, the peaks calculated with Eq. \eqref{eq: normalized power spectrum 2 cavities} are asymmetrical, resulting in small discrepancies from the sum of the Lorentzians, which are symmetrical by definition. 
The cause of this asymmetry is discussed in Appendix \ref{app: Asymmetric peaks}. In summary, we argue that these asymmetries should also occur in the PSDs of real stars. They thus take into account effects that are neglected when using a sum of Lorentz functions.
In any case, the deviations between our approach and the sum of Lorentzians are small (see Fig. \ref{fig: IRE_Lorentz_2cavities}) and irrelevant in the context of this work.

\subsection{Relative power spectrum} \label{sect: relative power spectrum}

As discussed in the previous section, the power spectrum $P$ is described by the normalized power spectrum $p$ up to a function that depends on the frequency-dependence of the excitation and damping in the p-mode cavity. Therefore, we can write the power spectrum as
\begin{gather}
    P(\nu) = f(\nu) \cdot p(\nu),
\end{gather}
where $f$ is a smooth function of $\nu$ that represents the influence of the excitation and damping in the p-mode cavity. Self-consistent consideration of these effects would go beyond the scope of this work, as it can be quite involved and requires numerical stellar models \citep[e.g.,][]{houdek+1999,dupret+09}. For clarity, we explicitly mention the frequency-dependence of the parameters in this and the following subsection (i.e., Sect. \ref{sect: visibility}).

Instead of modeling the excitation and damping processes in detail, we use the fact that the power spectrum $P$ follows a bell-shaped curve in the PSDs of solar-like oscillators. It is widely accepted that this curve can roughly be approximated by a Gaussian function with its center at the frequency of maximum oscillation power $\nu_{\rm max}$ \citep[e.g.,][]{mosser+12}. We thus assume that $f$ is proportional to a Gaussian function $G$ with with height equal to unity and standard deviation $\sigma_{\rm env}$. We then define the relative power spectrum $\mathcal{P}$ as the product of $p$ and the Gaussian function:
\begin{gather}
    \mathcal{P}(\nu) \equiv G(\nu) \cdot p(\nu).
    \label{eq: relative power spectrum}
\end{gather}
The standard deviation $\sigma_{\rm env}$ can be estimated using the empirical scaling-relation of \citet{mosser+12}, which depends on the stellar mass $M_*$ and $\Delta\nu$ (see Table \ref{tab: sets of stellar parameters}).
The relative power spectrum $\mathcal{P}$ is approximately proportional to the regular power spectrum $P$.

\subsection{Visibility} \label{sect: visibility}

The visibility $V^2_\ell$ is a proxy of the ratio between the average energy of the multipole modes of a given spherical degree $\ell$ and the radial modes. This can provide insights into the excitation and damping processes that take place in a star. The visibility can be calculated directly from a PSD as the ratio of the integrated power spectrum of the multipole modes to that of the radial modes over the range of observable frequencies, which typically ranges from $\nu_{\rm left} = \nu_{\rm max} - 3\ \Delta\nu$ to $\nu_{\rm right} = \nu_{\rm max} + 3\ \Delta\nu$:
\begin{gather}
    V^2_\ell = \frac{ \int_{\nu_{\rm left}}^{\nu_{\rm right}} P_\ell(\nu) \,\mathrm{d}\nu }{ \int_{\nu_{\rm left}}^{\nu_{\rm right}} P_{\ell=0}(\nu) \,\mathrm{d}\nu }.
\end{gather}
Using the proportionality of $P$ and the relative power spectrum $\mathcal{P}$ introduced in the previous section, we can express the visibility as
\begin{gather}
    V^2_\ell \approx \frac{ \int_{\nu_{\rm left}}^{\nu_{\rm right}} \mathcal{P}_\ell(\nu) \,\mathrm{d}\nu }{ \int_{\nu_{\rm left}}^{\nu_{\rm right}} \mathcal{P}_{\ell=0}(\nu) \,\mathrm{d}\nu }.
    \label{eq: visibility}
\end{gather}
Our framework is therefore suitable for calculating theoretical visibilities, as the proportionality constant connecting $\mathcal{P}$ to $P$ is eliminated in the fraction.
In this theoretical study, we neglect the additional $\ell$-dependent scaling factor affecting the observed visibility measurements, which is caused by geometric effects such as limb darkening and bolometric correction \citep[e.g.,][]{ballot+11}.

\subsection{A resolution criterion} \label{sect: a resolution criterion}

\begin{figure}[]
    
    \resizebox{\hsize}{!}{\includegraphics{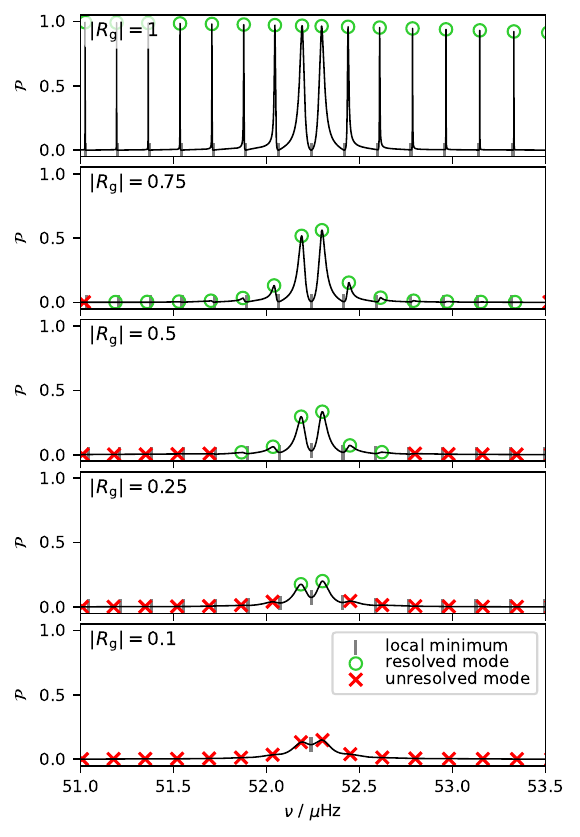}}
    \caption{Relative power spectrum as a function of frequency for the set of stellar parameters B with $|R_{\rm p}|=0.95$ and different values of $|R_{\rm g}|$. Gray vertical lines indicate the location of the local minima, green circles (red crosses) indicate the location of the resolved (unresolved) modes. The resolution criterion is described in Sect. \ref{sect: a resolution criterion}. When the unresolved modes no longer corresponded to local maxima on the selected frequency grid, we used the resonance condition to determine their position.}
    \label{fig: taylor criterion}
\end{figure}

As shown in Fig. \ref{fig: IRE_Lorentz_2cavities}, the number of recognizable peaks decreases with decreasing $|R_{\rm g}|$. This is particularly relevant for red giant stars, as there are examples with a low multipole mode visibility where mixed modes do not appear to be present in observations \citep[e.g.,][]{stello+16b}, and others that show clear signs of mixed modes \citep{mosser+17, arentoft+17}. We therefore investigate what conditions the damping processes in the g-mode cavity must satisfy in order for the mixed modes to be resolved and detectable in observations.

In optics, the Taylor criterion states that two peaks are resolved if they intersect at half their maximum height or below. Inspired by this, we first identify all local maxima and minima of the relative power spectrum $\mathcal{P}$. Every maximum is surrounded by two minima. If the value of $\mathcal{P}$ at a local maximum is greater than or equal to the values of $\mathcal{P}$ at the neighboring minima multiplied by 2, we consider the peak to be resolved. Otherwise, it is unresolved. We show an example of this procedure in Fig. \ref{fig: taylor criterion}.
In the rest of this work, we distinguish between two cases. If at least two peaks per pressure radial order are resolved according to our criterion, we assume that the mixed modes are theoretically detectable. If this is not the case, we assume that there is no observable signature of the mixed modes, which corresponds to the lower panel of Fig. \ref{fig: taylor criterion}.

Since we do not take into account any of the complications that might arise during observations, such as residuals from the background correction, stochastic excitation, and noise, our detection criterion can be considered a best-case scenario. This means that even though we might expect to detect mixed modes according to our resolution criterion, they may be obscured in observations. Conversely, however, we can make the solid prediction that we do not expect any signs of mixed modes in observations if there are no resolved peaks according to our resolution criterion.

\section{Results} \label{sect: Results}

\begin{table}[]
    \caption{Sets of stellar parameters used in this study.}
    \centering
    \begin{tabular}{c|ccc}
    \hline\hline
    model & A & B & C \\
    \hline
    evolution & early RGB & late RGB & CHeB \\
    $\nu_{\rm max}\ /\ \mu$Hz & 200 & 50 & 35 \\
    $\Delta\nu\ /\ \mu$Hz & 16 & 5.5 & 4.2 \\
    $\Delta\Pi_{\ell=1}$ / s & 85 & 65 & 310 \\
    $q$ & 0.125 & 0.04 & 0.2 \\
    $\sigma_{\rm env}\ /\ \mu$Hz & 22.6 & 7.8 & 5.9 \\
    \hline
    \end{tabular}
    \tablefoot{RGB stands for “red-giant branch star” and CHeB for “core-helium-burning star”. The selection of parameters is based on a numerical stellar evolutionary track with a mass of 1.25 $M_\odot$ and an approximate solar metallicity calculated with MESA \citep{mesa1}. The values of $q$ and $\Delta\Pi_{\ell=1}$ correspond to the dipole modes.}
    \label{tab: sets of stellar parameters}
\end{table}

In this section, we introduce some initial applications of the relative power spectrum $\mathcal{P}$. In Sect. \ref{sect: results setup}, we prepare the presentation of our results. In Sect. \ref{sect: results parameter study}, we perform a parameter study on visibility to understand how it behaves at finite damping rates in the g-mode cavity. In Sect. \ref{sect: results detectability of mixed modes}, we investigate the detectability of the mixed mode and the multiplet signature for the same parameter space, and in Sect. \ref{sect: results application mosser}, we apply our method to data from observations to infer the damping rate in the g-mode cavity of real red giant stars.
Note that in Appendix \ref{app: Parameters describing mode damping}, we summarize all parameters we use to describe the damping of the waves and their interrelations.

\subsection{Setup} \label{sect: results setup}

While the concepts discussed in the previous sections are quite general, we make a number of simplifying assumptions here in order to limit the scope of this work. First, as mentioned above, we assume that all reflection coefficients (i.e., $R_{\rm p}$, $R_q$, and $R_{\rm g}$) are constant and do not change with frequency. This assumption will be relaxed in a future study. Furthermore, for most of this work, we assume that $\epsilon_{\rm p} = \epsilon_{\rm g} = 0$, which physically corresponds to the Cowling approximation and the additional assumption that the boundaries of the cavities in a two-cavity system are far apart \citep[i.e., the weak coupling approximation;][]{pincon+19, pincon+takata22}. We examine the influence of $\epsilon_{\rm p}$ and $\epsilon_{\rm g}$ on the visibilities and the detectability of the mixed modes in Appendix \ref{app: Parameter study}.

In our calculation of $\Theta_{\rm p}$ in Eq. \eqref{eq: theta_p}, we neglected the $\ell$-dependence of $k_{r}$ because we assumed that $\omega^2 \gg N^2, S_\ell^2$ applies throughout the p-mode cavity. However, this assumption only holds deep inside the p-mode cavity and is not appropriate near the boundaries. To better reconcile $\Theta_{\rm p}$ with the observed universal pattern of acoustic modes, we follow previous studies \citep[e.g.,][]{mosser+12c,deheuvels+15,mosser+15} and introduce an additional term $\pi\ell/2$ into Eq. \eqref{eq: theta_p}, so that $\Theta_{\rm p}$ is instead given by \citep[see also equation (53) of][]{hekker+17}
\begin{gather}
    \Theta_{\rm p} = \pi \left(\frac{\nu}{\Delta\nu} + \frac{\ell}{2}\right).
    \label{eq: theta_p with ell}
\end{gather}
This additional contribution shifts the frequency of the p-modes, while their periodicity remains unchanged.

In addition to mixed mode singlets, which exhibit a single peak in the PSD per mixed mode, we also investigate the detectability of mixed mode multiplets, which exhibit multiple peaks per mixed mode. In this work, we assume that the physical process responsible for the splitting of the mixed mode into multiplets only operates in the g-mode cavity. Therefore, we introduce a frequency shift $\delta\nu$ to the phase $\Theta_{\rm g}$:
\begin{gather}
    \Theta_{\rm g} = -\frac{\pi}{\Delta\Pi_\ell\ (\nu +m\ \delta\nu)}.
    \label{eq: frequency shift theta_g}
\end{gather}
Here, $m$ is the azimuthal order, which is an integer bounded by $-\ell \leq m \leq \ell$. In red giant stars, symmetric frequency shifts similar to the one described in Eq. \eqref{eq: frequency shift theta_g} can be induced by, for example, rotation \citep[e.g.,][]{beck+2012, mosser+12, deheuvels+12}. If $\delta\nu \neq 0$, we calculate $\mathcal{P}$ of a given $\ell$ separately for all values of $m$. We then sum the corresponding $\mathcal{P}$ of each $m$ to obtain the total relative power spectrum of this spherical degree. In this procedure, we take into account the $m$-dependent scaling factors of the power spectrum \citep[e.g.,][]{gizon&solanki03}. These scaling factors depend on the inclination angle $i$ of the star. Unless specified otherwise, we use $i = 0^\circ$ (i.e., mixed mode singlets).

Finally, when showing our results, we use three sets of stellar parameters that roughly represent different stages in the evolution of a red giant star with a mass of 1.25 $M_\odot$ and an approximate solar metallicity. The values corresponding to each set are listed in Table \ref{tab: sets of stellar parameters}. 
In this study, we restrict ourselves to the radial ($\ell=0$) and dipole ($\ell=1$) modes. Since $q$ and $\Delta\Pi_\ell$ depend on $\ell$, their estimates in Table \ref{tab: sets of stellar parameters} refer to the dipole modes\footnote{For the radial modes, we use $q=0$ and $\Delta\Pi_{\ell=0}=\infty$.}.
All changes to the default parameters specified in Table \ref{tab: sets of stellar parameters} are indicated in the figure captions.

\subsection{Parameter study on the visibility} \label{sect: results parameter study}

\begin{figure*}[]
    
    \resizebox{\hsize}{!}{\includegraphics{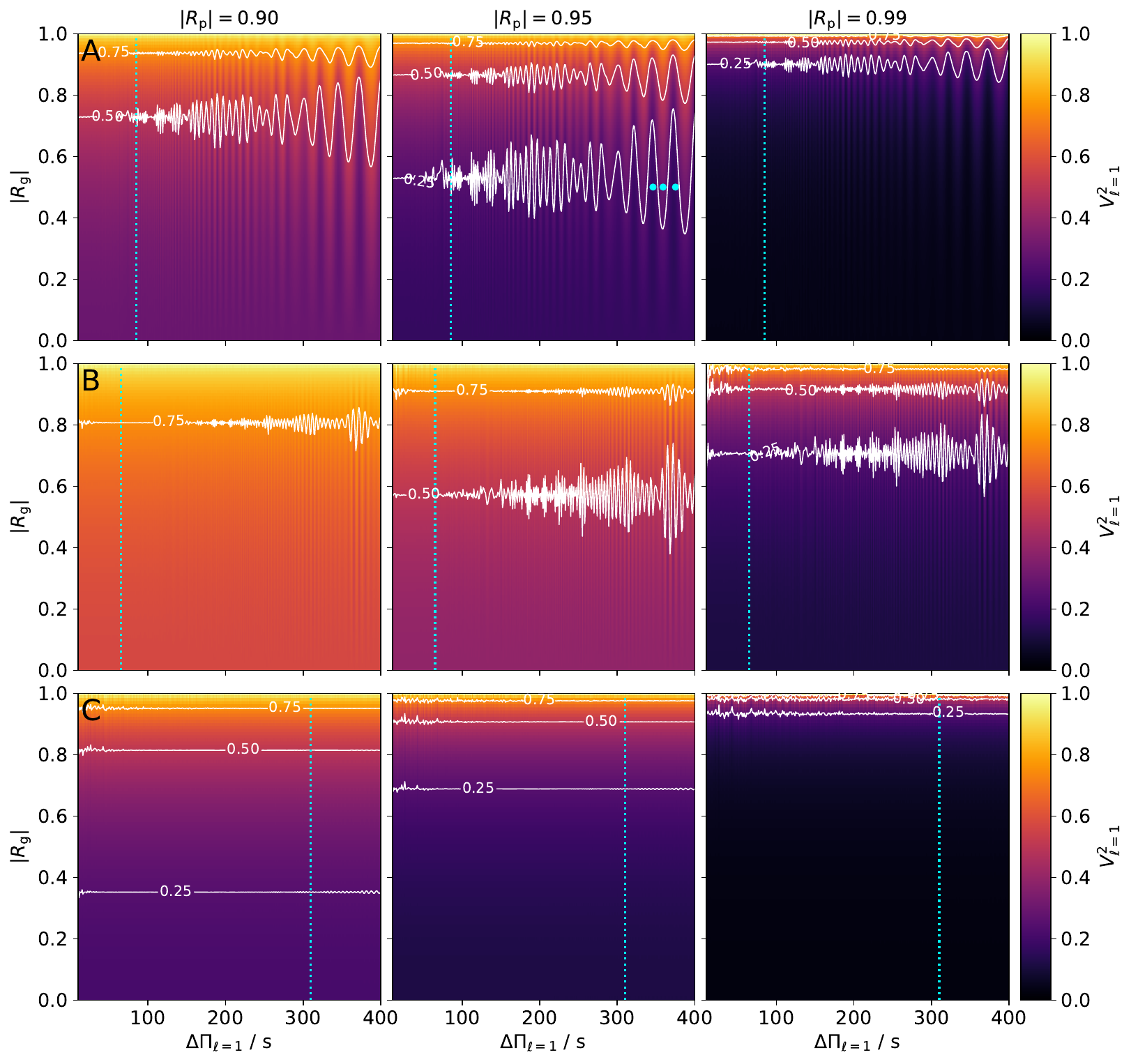}}
    \caption{Dipole mode visibility as a function of period spacing and the modulus of the reflection coefficient of the lower boundary of the g-mode cavity. {\it Rows} correspond to different sets of stellar parameters (see Table \ref{tab: sets of stellar parameters}), {\it columns} correspond to different values of the reflection coefficient of the upper boundary of the p-mode cavity. White contour lines mark bands with equal visibility (see labels). Cyan dotted lines mark the nominal period spacing of the corresponding set of stellar parameters. Cyan dots indicate the parameter sets that are further examined in Fig. \ref{fig: ridges delta Pi}.}
    \label{fig: 3x3 visibility}
\end{figure*}
\begin{figure}[]
    
    \resizebox{\hsize}{!}{\includegraphics{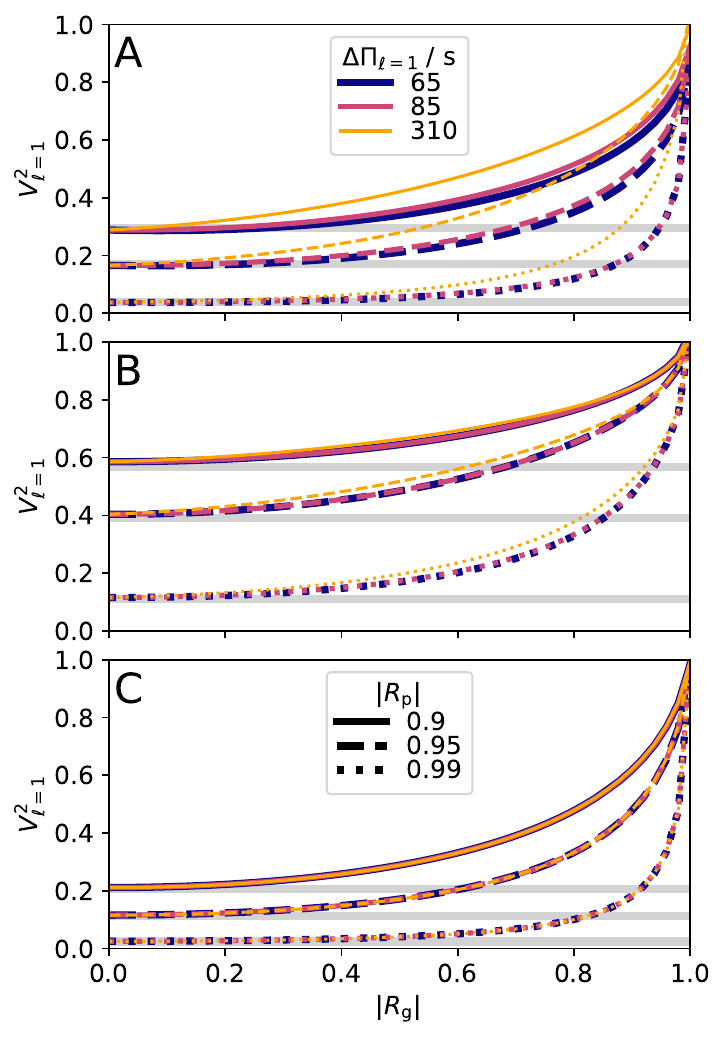}}
    \caption{Dipole mode visibility as a function of the modulus of the reflection coefficient of the lower boundary of the g-mode cavity for our three sets of stellar parameters. For each set, we show the visibility for the three nominal values of the period spacing of the parameters sets (colors) and the modulus of the reflection coefficient of the upper boundary of the p-mode cavity (line styles). The gray horizontal lines correspond to the visibility of the dipole mode when $R_{\rm g}=0$ (Eq. \eqref{eq: visibility full suppression}). Small deviations from the value predicted by Eq. \eqref{eq: visibility full suppression} can be induced by the multiplication of the normalized power spectrum with the Gaussian envelope (see Sect. \ref{sect: relative power spectrum}). For an explanation why the visibility is not always equal to one when $|R_{\rm g}|=1$, see Appendix \ref{app: Variation in visibility no core damping}.}
    \label{fig: visibility versus R_g}
\end{figure}

For red giant stars, whose oscillations are reminiscent of the two-cavity system discussed above, it has been argued that the visibility of the multipole modes should be always equal to one if there is no energy loss in the g-mode cavity \citep{mosser+17}. This corresponds to $|R_{\rm g}| = 1$. In the other extreme case, $R_{\rm g} = 0$, in which all energy entering the g-mode cavity is lost, the visibility is given by the ratio of the damping rate of a radial mode to that of a multipole mode \citep{takata16b, mosser+17}:
\begin{gather}
    V^2_\ell = -\frac{\eta_{\rm p}}{2\pi\nu_\Im} = \left(1 - \ln\left(1 - \frac{4q}{(1+q)^2}\right) \frac{\Delta\nu }{2 \eta_{\rm p}} \right)^{-1}.
    \label{eq: visibility full suppression}
\end{gather}
This expression can be derived by substituting $\nu_\Im$ in Eq. \eqref{eq: visibility full suppression} using Eq. \eqref{eq: resonance condition 1 cavity imag}, since this scenario is equivalent to a single cavity with energy losses on both sides (see also Appendix \ref{app: Parameters describing mode damping}). If $0 < |R_{\rm g}| < 1$, there is currently no such expression for the visibility in the literature. We therefore use the method described in Sect. \ref{sect: Synthetic power spectrum} to estimate the visibility of the dipole modes for different values of $|R_{\rm p}|$, $|R_{\rm g}|$, and $\Delta\Pi_{\ell=1}$. The results for our three sets of stellar parameters are shown in Fig. \ref{fig: 3x3 visibility}. In Appendix \ref{app: Parameter study}, we also investigate different values for $q$, $\epsilon_{\rm p}$, and $\epsilon_{\rm g}$.

A general trend that can be observed in all panels of Fig. \ref{fig: 3x3 visibility} is that the visibility decreases with decreasing $|R_{\rm g}|$. This is to be expected, since smaller values of $|R_{\rm g}|$ mean that more oscillation energy is lost in the g-mode cavity, hence reducing the observable average energy of the mixed modes in the p-mode cavity. In fact, $V^2_{\ell=1} \approx 1$ when $|R_{\rm g}| = 1$. 
As $|R_{\rm g}|$ decreases, the visibility approaches the value predicted by Eq. \eqref{eq: visibility full suppression}, seamlessly connecting the two extremes.
This is illustrated in more detail in Fig. \ref{fig: visibility versus R_g}.

Interestingly, the visibility is not exactly equal to one in most cases where there is no damping in the g-mode cavity. This differs from an analytical argument by \citet{mosser+17} and is a consequence of the fact that the simplifying assumptions they made in the reasoning in their section 2.2 are not always entirely valid. We discuss this in more detail in Appendix \ref{app: Variation in visibility no core damping}. Nevertheless, the deviations are so small that, for most practical purposes, the assumption that the visibility is equal to one when there is no damping in the g-mode cavity is justified.

Another trend that is clearly visible in Fig. \ref{fig: 3x3 visibility} is that the visibility is generally lower for larger values of $|R_{\rm p}|$ when $|R_{\rm g}| < 1$. This corresponds to the behavior predicted by Eq. \eqref{eq: visibility full suppression}. The physical reason for this is that the visibility is the ratio of the integrated power spectrum of the multipole modes to that of the radial modes. While the multipole modes are damped in both the p- and g-mode cavities, the radial modes are only affected by the damping in the p-mode cavity. For large values of $|R_{\rm p}|$, the radial modes lose only a small amount of energy, making the visibility more sensitive to the energy loss of the multipole modes in the g-mode cavity. For smaller values of $|R_{\rm p}|$, the radial modes are more strongly damped and their integrated power spectrum is lower, which increases the visibilities.

Furthermore, the visibilities corresponding to the set of stellar parameters B are generally higher than those of sets A and C. This is consistent with Eq. \eqref{eq: visibility full suppression}. The main reason for this difference is that set B has a much lower coupling strength $q$ than the other sets. Since $q$ measures the interaction between the p- and g-mode cavities, a smaller value of $q$ means that less energy enters the g-mode cavity (see Eq. \eqref{eq: coupling strength}). Thus, the multipole modes are less affected by energy loss in the g-mode cavity, while the radial modes remain unaffected, which increases visibility. The influence of $q$ on visibility is examined in more detail in Appendix \ref{app: Parameter study}.

The visibilities shown in Fig. \ref{fig: 3x3 visibility} increase and decrease alternately with varying $\Delta\Pi_{\ell=1}$. This is reflected in Fig. \ref{fig: 3x3 visibility} in the form of vertical ridges, which lead to an apparent oscillatory behavior of the contour lines. The width of these ridges decreases with increasing number of mixed modes in a pressure radial order (i.e., with increasing number of modes with the same integer $n$ in Eq. \eqref{eq: resonance condition 2 cavities with n}). Therefore, the ridges become smaller and eventually disappear at smaller values of $\Delta\Pi_{\ell=1}$ and $\nu_{\rm max}$. To investigate this more closely, we consider the power spectrum of the parameters corresponding to the cyan dots in Fig. \ref{fig: 3x3 visibility} further in Appendix \ref{app: Additional figures}.
In short, the visibility of mixed modes is higher when the frequency of the nominal p-modes is close to that of a nominal g-mode.

It is also interesting to note that Eq. \eqref{eq: visibility full suppression} does not necessarily give the minimum value of the visibility for $0 \leq |R_{\rm g}| \leq 1$. Due to the ridges mentioned in the previous paragraph, it is possible that the frequencies of the nominal p- and g-modes are so far apart from each other that the visibility is even lower than if no energy were returning from the g-mode cavity. The difference is usually only a few percent and is therefore probably too small to be detected in observations. In fact, Fig. \ref{fig: visibility versus R_g} shows that $V^2_{\ell=1}$ generally resembles the value predicted by Eq. \eqref{eq: visibility full suppression} within a few percent for a range of $|R_{\rm g}|$ that depends on the choice of the other parameters. We can therefore conclude that in observations, the visibility is no longer distinguishable from the value given by Eq. \eqref{eq: visibility full suppression} once $|R_{\rm g}|$ has become smaller than a threshold value that depends on the parameters of the star under consideration. The uncertainty of $q$ also contributes to this, as it can only be constrained in observations if a sufficient number of mixed modes can be identified in the PSD.

Finally, we point out that although we focus on dipole modes in this study, the general trends in visibility discussed here also apply to multipole modes with a higher spherical degree. However, the values for period separation and coupling strength must be adjusted accordingly. This will be addressed in a future study.

\subsection{Detectability of the mixed modes} \label{sect: results detectability of mixed modes}

\begin{figure*}[]
    
    \resizebox{\hsize}{!}{\includegraphics{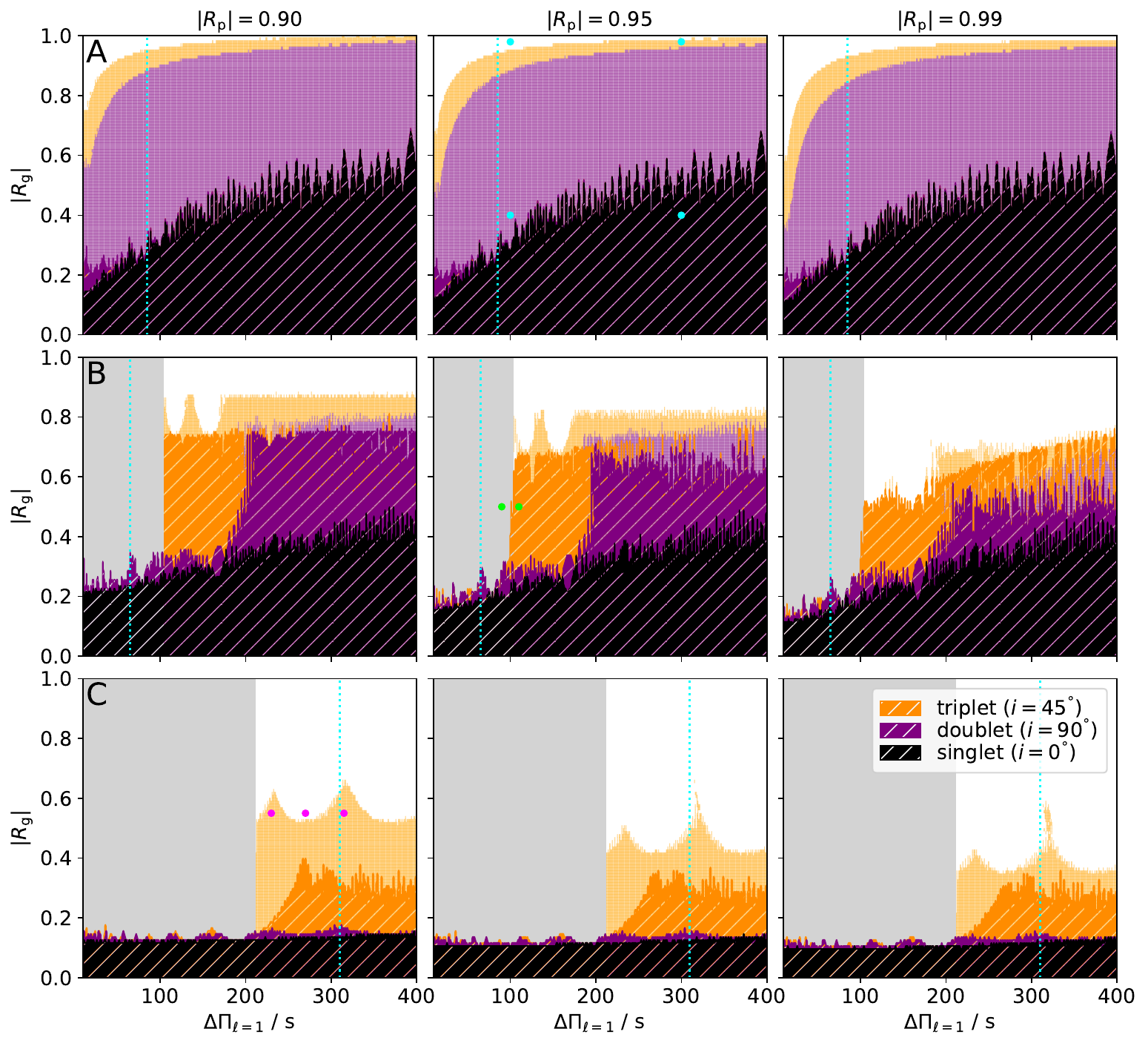}}
    \caption{Detectability of the dipole modes as a function of period spacing and the modulus of the reflection coefficient of the lower boundary of the g-mode cavity for a frequency shift of $\delta\nu = 0.1\ \mu$Hz. {\it Rows} correspond to different sets of stellar parameters, {\it columns} correspond to different values of the reflection coefficient of the upper boundary of the p-mode cavity.
    White areas indicate regions where both the mixed mode and multipole signatures can be detected for all tested inclination angles.
    Hatched areas indicate combinations for which mixed modes cannot be detected. 
    Lightly colored areas indicate combinations of parameters for which the multiplet signature cannot be detected. 
    Gray areas indicate regions where the average frequency separation of the mixed modes becomes comparable to $\delta\nu$.
    Other colors correspond to different inclination angles of the star. Black areas usually overlap with purple areas, and purple areas usually overlap with orange areas.
    Cyan dotted lines mark the nominal period spacing of the corresponding set of stellar parameters. Cyan dots indicate the sets of parameters that are further investigated in Fig. \ref{fig: resolution delta Pi 2}, green dots mark the parameter sets shown in Fig. \ref{fig: resolution delta Pi}, and magenta dots mark the sets shown in Fig. \ref{fig: resolution delta Pi 3}.}
    \label{fig: 3x3 mixed mode detectability}
\end{figure*}

Since the lifetime $\tau$ of the oscillation modes is directly related to the linewidth of their corresponding peaks in the PSD, the mixed modes become broader as $|R_{\rm p}|$ and $|R_{\rm g}|$ decrease. Moreover, the additional energy loss reduces their peak height. This means that with sufficiently efficient damping, the signature of the mixed mode becomes smudged, as the individual peaks broaden so much that only low-amplitude clumps remain around the frequency of the nominal p-modes. It is likely that this case cannot be distinguished from pure p-mode oscillations in observations and thus mimics the case $R_{\rm g} = 0$.

Similarly, the individual peaks of mixed mode multiplets become broader as they lose more of their energy, eventually causing the split signature of the multiplets to disappear. Since the multiplet signature is crucial for asteroseismic techniques used to measure both rotational and magnetic perturbations to the restoring forces of the wave \citep[e.g.,][]{beck+2012,li+22}, these techniques can no longer be used when the damping in the g-mode cavity suppresses the presence of the multiplets in observations. For simplicity, we restrict ourselves to symmetric dipole mode multiplets and investigate the detectability of the split peak for a frequency shift of $\delta\nu = 0.1\ \mu$Hz. This is roughly reminiscent of the signature induced by a rotating core in red giant stars.
Since the number of peaks of a multiplet visible in the PSD depends on the inclination angle $i$ of the star, we investigate the detectability of the multiplets for three different inclination angles ($i = 0^{\circ},\ 45^{\circ}$, and $90^{\circ}$, corresponding to mixed mode singlets, triplets, and doublets, respectively).

In Fig. \ref{fig: 3x3 mixed mode detectability}, we show the detectability of mixed modes according to the resolution criterion discussed in Sect. \ref{sect: a resolution criterion} for the same parameter space as shown in Fig. \ref{fig: 3x3 visibility}. We also show the detectability of the split nature of the multiplets. To do this, we use the same resolution criterion as for the signature of the mixed modes. However, instead of finding at least two resolved modes in a frequency range of one pressure radial order, we look for the presence of two or more peaks in a range of width $2 \delta\nu$ centered around each peak in the PSD corresponding to the singlets (i.e., $i=0^\circ$). In other words, we look for additional peaks in a frequency range around each unsplit peak. If we find at least two peaks in the range around one or more unsplit peaks, we consider the multiplets to be theoretically detectable.
However, this also means that we detect false positives when the frequency separation between the mixed modes approaches $\delta\nu$. We determined the parameter range in which the mixed modes and multiplets become ambiguous by searching for multiplets in the relative power spectrum corresponding to an inclination angle of $i=0^\circ$ (in which only mixed mode singlets are present) and marked it in gray in Fig. \ref{fig: 3x3 mixed mode detectability}. The gray area appears vertically because $\Delta\Pi_{\ell=1}$ must lie below a certain threshold such that the number of the mixed modes is high enough for this to be the case.

Although Fig. \ref{fig: 3x3 mixed mode detectability} shows a variety of trends that are difficult to summarize in a few sentences, there are three general trends that can be recognized in each panel. The first trend is that, for a given inclination angle $i$, the multiplet signature always disappears before or simultaneously with the mixed mode signature. The reason for this is that the criterion for detecting multiplets is more stringent than that for detecting mixed modes, since multiplets require at least two detectable peaks in a frequency range of a width $2\delta\nu$, whereas we consider a width of $\Delta\nu$ for the detectability of mixed modes.

The second global trend is that, for both the multiplet and mixed mode signatures, the parameter space in which they are detectable is smallest for the triplets and largest for the singlets, while that for the doublets lies in between. The reason for this is quite illustrative. The greater the number of peaks in a multiplet, the smaller their individual maximum height, since the total amplitude of a multiplet remains constant as a function of $i$. As mentioned earlier, the more energy is lost in the g-mode cavity, the broader the mode peaks become. Both the higher number of peaks and their lower height contribute to the smudging of the peaks with decreasing $|R_{\rm g}|$, as they become wider and thus reduce the detectability of the individual peaks.

The third global trend is that the dependence of the region in which the multipole modes are not detectable when $i=0^\circ$ (i.e., singlets) becomes flatter from set A to B to C. The reason for this lies in the different number of mixed modes corresponding to the sets. Since $\nu_{\rm max}$ decreases from set A to C, the number of mixed modes increases. For the value of $\nu_{\rm max}$ corresponding to set C, even at $\Delta\Pi_{\ell=1} = 400$ s there are so many mixed modes that their exact number is no longer relevant for our detectability criterion of the mixed modes at $i=0^\circ$. They clump at approximately the same value of $|R_{\rm g}|$, causing the threshold value of the range in which the mixed mode singlets are no longer detectable to appear as an almost horizontal line in lower row of Fig. \ref{fig: 3x3 mixed mode detectability}.

We discuss a selection of additional, more specific trends in Appendix \ref{app: Additional figures} using the colored dot families in Fig. \ref{fig: 3x3 mixed mode detectability}. Overall, we conclude that energy loss in the core can conceal both the mixed mode and the multiplet signature in the PSD if the damping in the g-mode cavity is sufficiently efficient. In particular, it is therefore possible that stars with a reflection coefficient of $|R_{\rm g}|>0$ appear as if $|R_{\rm g}|$ were exactly 0, with no mixed modes present. The exact threshold value of $|R_{\rm g}|$ at which the signatures disappear depends strongly on stellar parameters such as $|R_{\rm p}|$, $\Delta\nu$, $\Delta\Pi_{\ell=1}$, and $i$, as well as on $q$, $\epsilon_{\rm p}$ and $\epsilon_{\rm g}$ (see Appendix \ref{app: Parameter study}). The detectability of these features must therefore be assessed on a case-by-case basis if robust quantitative results are required.

\subsection{Application to the sample of \citet{mosser+17}} \label{sect: results application mosser}

\begin{figure}[]
    
    \resizebox{\hsize}{!}{\includegraphics{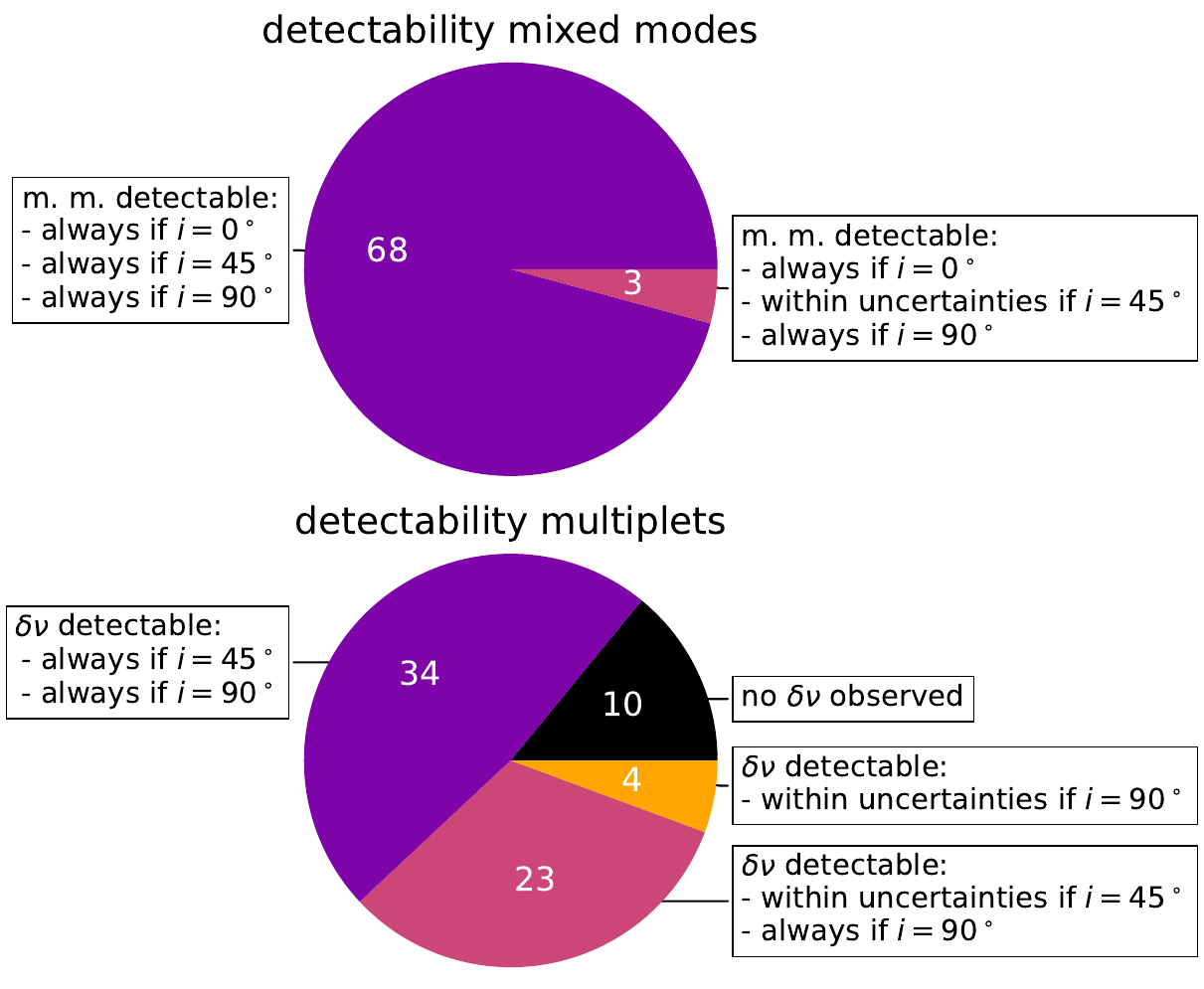}}
    \caption{Pie chart showing the detectability of mixed mode (m. m.) and multiplet signatures according to our resolution criterion for 71 red giant stars using the observationally constrained stellar parameters from \citet{mosser+17}. The detectability depends on the inclination angle $i$. “Always detectable” means that the feature is resolved for all tested values of radial mode linewidth (i.e., $0.7\ \Gamma_0$, $\Gamma_0$, and $1.3\ \Gamma_0$). “Detectable within the uncertainties” means that the feature is resolved for at least one value of a tested radial mode linewidth. “No $\delta\nu$ observed” corresponds to the 10 stars for which \citet{mosser+17} did not measure a frequency shift.}
    \label{fig: piechart_mosser17}
\end{figure}
\begin{figure}[]
    
    \resizebox{\hsize}{!}{\includegraphics{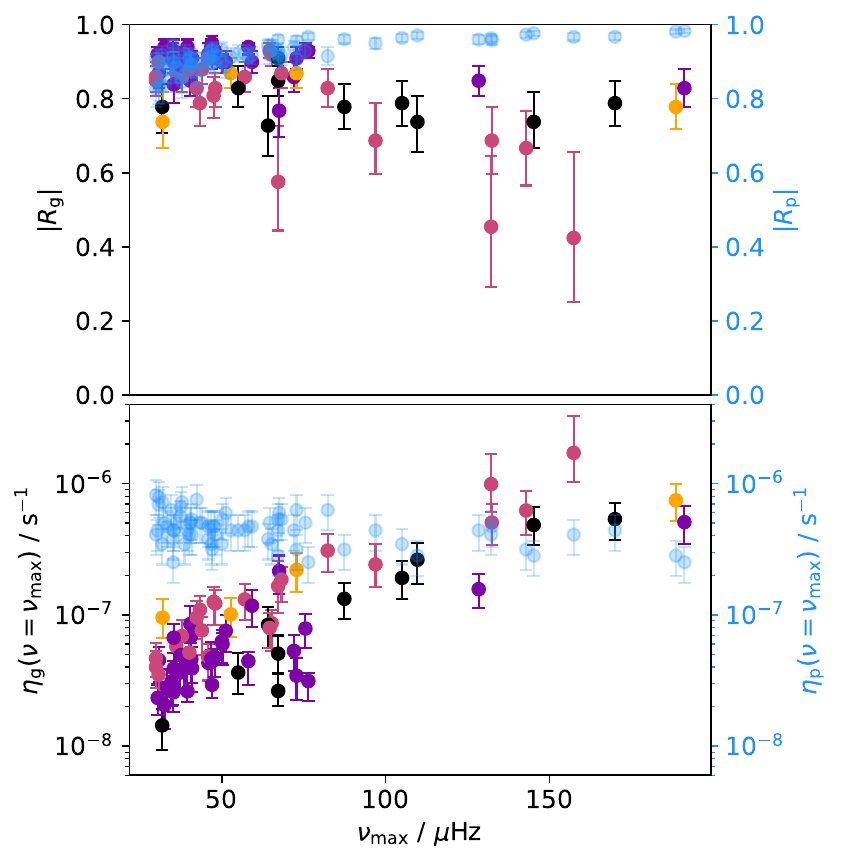}}
    \caption{Modulus of the reflection coefficient at the lower boundary of the g-mode cavity {\it (upper panel)} and damping rate in the g-mode cavity at the frequency $\nu_{\rm max}$ {\it (lower panel)} as a function of frequency of maximum oscillation power for the sample of red giants of \citet{mosser+17}. The same quantities are shown for the upper boundary of the p-mode cavity in transparent blue. The colors of the solid symbols correspond to the lower pie chart in Fig. \ref{fig: piechart_mosser17} (i.e., to the detectability of the multiplet signature). Errors correspond to the uncertainties in the observed radial mode linewidth (see main text).}
    \label{fig: parameters_mosser17}
\end{figure}

In their observational study, \citet{mosser+17} identified 71 red giant stars with a low dipole mode visibility that show clear signs of dipolar mixed modes. For these stars, the authors measured the dipole mode visibility $V^2_{\ell=1}$ and other stellar parameters such as $\nu_{\rm max}$, $\Delta\nu$, $\Delta\Pi_{\ell=1}$, $q$, and $M_*$. In addition, they measured the average radial mode linewidth $\Gamma_0$ with an uncertainty of 30\%, which is related to the damping rate in the p-mode cavity via $\eta_{\rm p} = \pi\Gamma_0$ \citep[e.g.,][]{samadi+15}.
The damping rate, in turn, is related to $\mu_{\rm p}$ due to Eq. \eqref{eq: imaginary nu pressure} (see also Appendix \ref{app: Parameters describing mode damping}). For 61 of these stars, \citet{mosser+17} were also able to constrain the rotation-induced frequency shift $\delta\nu$. In fact, they measured all the input parameters required for our approach to calculating the synthetic PSD (excluding the phase shifts $\epsilon_{\rm p}$ and $\epsilon_{\rm g}$), with the exception of $\mu_{\rm g}$.

However, since \citet{mosser+17} provided an estimate for $V^2_{\ell=1}$\footnote{Note that in observed PSDs, it is not feasible to completely separate the contribution of the peaks corresponding to different spherical degrees, as we are doing here. It is therefore possible that the spread of the gravity-dominated dipole modes away from the location of the nominal p-mode leads to them being neglected in the calculation of dipole mode visibility, thus introducing a bias \citep[see][]{mosser+12}. Taking this circumstance into account in our theoretical estimates would go beyond the scope of this work and will be addressed in a future study. Nonetheless, this should not strongly affect the results presented here, since the gravity-dominated modes are the first to disappear from the PSD when energy losses occur in the g-mode cavity (see Figs. \ref{fig: IRE_Lorentz_2cavities} and \ref{fig: taylor criterion}).}, we can constrain $|R_{\rm g}|$ (and thus $\mu_{\rm g}$) as follows. We generate estimates of the dipole mode visibility for 100 realizations of the synthetic PSD for each star of their sample. To compute these realizations, we use the same value for $M_*$, $\nu_{\rm max}$, $\Delta\nu$, $\Delta\Pi_{\ell=1}$, $q$, $\mu_{\rm p}$ and $\delta\nu$ as given by \citet{mosser+17}, but vary the value of $|R_{\rm g}|$ on a uniform grid from 0 to 1. This gives us a different estimate of $V^2_{\ell=1}$ for each value of $|R_{\rm g}|$, and we can subsequently select the value of $|R_{\rm g}|$ for which $V^2_{\ell=1}$ most closely matches the observed value given by \citet{mosser+17}.
Since $V^2_{\ell=1}$ does not always change monotonically with $|R_{\rm g}|$, we have ensured that the observed value of $V^2_{\ell=1}$ can be unambiguously assigned to a value of $|R_{\rm g}|$ for each of the 71 stars.
We then repeat the same process twice for different values of $\Gamma_0$, namely $0.7\ \Gamma_0$ and $1.3\ \Gamma_0$, in order to quantify the uncertainty of our estimate of $|R_{\rm g}|$.
The uncertainties can be estimated in this way, since we tested that $V^2_{\ell=1}$ does change monotonically with the linewidth $\Gamma$ when $0.7\ \Gamma_0 \leq \Gamma \leq 1.3\ \Gamma_0$.
We assume here that the uncertainty of the observed value of the average radial mode linewidth outweighs the uncertainties of the other stellar parameters, so that these can be neglected.
With the newly obtained values of $|R_{\rm g}|$, we can generate synthetic PSDs for all 71 stars in the sample of \citet{mosser+17}. In particular, this means that we can apply our resolution criterion to both the mixed mode signatures and the multiplet signatures and compare our predictions of the detectability of the modes with the observations. We investigate the detectability of the multiplets using the observed value of $\delta\nu$ for the inclination angles $i = 45^\circ$ and 90$^\circ$. For the detectability of the mixed modes, we also consider $i = 0^\circ$.

Using the reflection coefficients $|R_{\rm p}|$ and $|R_{\rm g}|$, we determine the damping rates in the p- and g-mode cavities (i.e., $\eta_{\rm p}$ and $\eta_{\rm g}$). Unlike in Sect. \ref{sect: pressure and gravity modes}, where we considered idealized oscillation modes with complex eigenfrequencies, we assume here that $\nu$ is real. Therefore, we use Eqs. \eqref{eq: crossing time} and \eqref{eq: amplitdue modification vs. eta} to derive the corresponding expressions for $\eta_{\rm p}$ and $\eta_{\rm g}$ instead of using Eqs. \eqref{eq: imaginary nu pressure} and \eqref{eq: imaginary nu gravity}. For the p-mode cavity, we recover Eq. \eqref{eq: imaginary nu pressure}. For the g-mode cavity, we must replace $|\nu|^2$ with $\nu^2$ in Eq. \eqref{eq: imaginary nu gravity}, where $\nu$ is a real number. In summary, we use
\begin{gather}
    \eta_{\rm p} = 2\ \Delta\nu\ \mu_{\rm p} \quad\text{and}\quad \eta_{\rm g} = 2\ \Delta\Pi_\ell\ \nu^2\ \mu_{\rm g}
\end{gather}
to calculate the damping rates (see also Appendix \ref{app: Parameters describing mode damping}). We also compute the uncertainties of the damping rates from the uncertainties of $\mu_{\rm p}$ and $\mu_{\rm g}$.

The results of this analysis are shown in Figs. \ref{fig: piechart_mosser17} and \ref{fig: parameters_mosser17}. Overall, we find excellent agreement between our predictions for the detectability of the mixed mode and multiplet signatures and the observations of \citet{mosser+17}. Looking at Fig. \ref{fig: piechart_mosser17}, we find that mixed modes are detectable for all tested parameter combinations for 68 of the 71 stars. For the remaining three stars, this is only possible if $i=0^\circ$ or 90$^\circ$. However, the detection of the mixed mode signature is still possible even if $i = 45^\circ$ for values of $\Gamma_0$ within the observational uncertainties. Overall, our detection criterion for the mixed mode signature seems to work as expected.

A similar picture emerges for the detectability of the reported frequency shifts $\delta\nu$. Of the 61 stars for which $\delta\nu$ was measured, the multiplets of about half are detectable for all tested parameter combinations. For about a quarter, they are detectable for all tested parameter combinations when $i = 90^\circ$, while this is only possible for some values of $\Gamma_0$ within the observed uncertainties when $i = 45^\circ$. For four stars, the multiplets are theoretically only detectable when $i = 90^\circ$ and only for some values of $\Gamma_0$ within the observed uncertainties. Nevertheless, the detection of these frequency shifts is still possible under the right conditions according to our resolution criterion.
We therefore conclude from Fig. \ref{fig: piechart_mosser17} that there is not a single star for which the observed values of \citet{mosser+17} contradict our theoretical predictions. This is an encouraging result that supports the applicability of the method presented in this work.

In Fig. \ref{fig: parameters_mosser17}, we show our estimates for $|R_{\rm g}|$ and $\eta_{\rm g}$ at $\nu = \nu_{\rm max}$ together with the same quantities for the damping in the p-mode cavity. From that figure, we can deduce that the damping in the g-mode cavity appears to become less effective as the star evolves (i.e., as $\nu_{\rm max}$ decreases). In particular, the damping in the g-mode cavity tends to be stronger than in the p-mode cavity on the early red-giant branch, while it appears to be significantly weaker on the late red-giant branch and in the core-helium-burning stage.
In this context, it is important to note that the sample from \citet{mosser+17} is not representative of either the entire population of red giant stars or stars with unusually low multipole mode amplitudes, as only stars for which dipolar mixed modes could be identified were considered. In other words, the sample is strongly biased, as it consists only of red giant stars with a strong damping process in the g-mode cavity, which is however not efficient enough to suppress the mixed signature of the multipole modes. Therefore, the damping rates shown in Fig. \ref{fig: parameters_mosser17} are not representative of the evolution of the damping rate of red giant stars.

From Fig. \ref{fig: 3x3 mixed mode detectability}, we would expect the threshold value of $|R_{\rm g}|$ at which the mixed mode signature disappears to change little during evolution when a realistic estimate of $\Delta\Pi_{\ell=1}$ is considered. The expected value of $\Delta\Pi_{\ell=1}$ for a given set of stellar parameters is shown in Fig. \ref{fig: 3x3 mixed mode detectability} by the cyan dotted lines. This means that we would expect the mixed mode signature to be observable up to approximately the same value of $|R_{\rm g}|$ throughout the entire evolution shown in Fig. \ref{fig: parameters_mosser17}.
One possible reason for the trend in Fig. \ref{fig: parameters_mosser17} could be that the mixed mode signature required to determine the stellar parameters becomes increasingly difficult to fit. This is due to the larger number of mixed modes at lower $\nu_{\rm max}$ and the weaker coupling strength on the late red-giant branch, which makes the dipolar modes appear less mixed.
Furthermore, observational effects, such as residuals from background correction, stochastic excitation, and noise, which are not taken into account in our theoretical approach, could play a role.
In a future study, we will investigate the detectability of mixed modes and multiplet signatures along the evolution of the star using numerical stellar models. This could provide further insight into the trends visible in Fig. \ref{fig: parameters_mosser17}.

\section{Conclusions} \label{sect: Summary}

In this study, we have adopted the progressive wave picture discussed by \citet{takata16b} and \citet{pincon+takata22} to derive an analytical function that expresses the theoretical resonance pattern of solar-like oscillators and the amplitudes of their modes up to a proportionality factor. When applied to red giant stars, this function can model the effect of damping on the oscillations in both the p- and g-mode cavities, as well as the asymmetries of the mixed modes resulting from the interaction between the two cavities. Furthermore, it is possible to obtain an estimate for the visibility of the mixed modes, which is of interest due to the discovery of a subgroup of red giant stars with unusually low amplitudes \citep{mosser+12}.
The method presented in this study requires only a handful of input stellar parameters, which can be constrained either by observations or by numerical stellar models, and links arbitrary values of the theoretical damping rates in both cavities to the visibility, which can be constrained by observations.

First, we conducted a parameter study on visibility. We found that as energy loss in the g-mode cavity increases, the visibility steadily transitions between the two limits discussed in the literature from approximately one (for $|R_{\rm g}|=1$, see Appendix \ref{app: Variation in visibility no core damping}) to the value predicted by Eq. \eqref{eq: visibility full suppression} (for $R_{\rm g}=0$), which corresponds to complete energy loss in the g-mode cavity. In the latter case, visibility depends only on the coupling strength and damping in the p-mode cavity. When $|R_{\rm g}| > 0$, we find that, in addition to these two quantities, the visibility also depends on the period spacing and the phase shift induced onto the wave in the two cavities (i.e., $\epsilon_{\rm p}$ and $\epsilon_{\rm g}$, see Appendix \ref{app: Parameter study}). Furthermore, we find that as energy loss in the g-mode cavity increases, the visibility can approach the value predicted by Eq. \eqref{eq: visibility full suppression} quite rapidly. This means that the visibility of a red giant star with a strong damping mechanism in its core may be indistinguishable from that of a star in which all the energy entering the g-mode cavity is completely lost.

Apart from visibility, we find that a damping mechanism operating in the g-mode cavity affects the detectability of the mixed character of the multipole modes, as well as the multiplet signature induced by, for example, the rotation of the stellar core. While it is to be expected that these signatures disappear when all the energy entering the g-mode cavity is lost, our work shows that this can also happen when part of the energy returns from the g- to the p-mode cavity. The detection threshold depends on the individual stellar parameters and the inclination angle of the star, which affects the number of peaks observable for each mixed mode multiplet. As a rule of thumb, both the mixed mode and multiplet signatures are easier to detect the smaller the number of modes in a multiplet. In general, however, we recommend comparing the theoretical estimates and observations for each star individually, as the detection threshold for these signatures is sensitive to the parameters of a given star.

Finally, we applied our theoretical framework to each star in the sample of red giant stars presented by \citet{mosser+17}, which consists of stars with a low dipole mode visibility but which nevertheless exhibit mixed mode signatures in the dipole modes. For most of these stars, the authors were even able to measure the rotation-induced frequency shift based on the observed multiplet signature of the modes. 
We find that our detection criterion for both the mixed mode and the multiplet signatures agrees perfectly with the observations of \citet{mosser+17}. Furthermore, we were able to constrain the modulus of the reflection coefficient of the lower boundary of the g-mode cavity and the corresponding damping rate from the observed dipole mode visibilities of these stars. This demonstrates the potential of the method presented here, which links theoretical damping rates with observable visibilities.

Considering the results presented in this work, we can interpret the characteristics of the observed population of red giants with low multipole mode visibility of which some are consistent with complete energy dissipation in the g-mode cavity, while others are not \citep[e.g.,] []{mosser+12, stello+16a, stello+16b, mosser+17}.
In each of these stars, a strong damping process operates in the core, reducing the amplitudes of their multipole modes \citep[e.g.,][]{garcia+14, fuller+15, cantiello+16, coppee+24}. If we assume that the corresponding damping rate is finite, a small fraction of the mode energy can escape and interact with the p-mode cavity. This would mean that the multipole modes with low amplitudes are intrinsically mixed. However, depending on the individual properties of the stars and the strength of the damping rate in the g-mode cavity, the signature of the mixed modes may not be detectable in observations, so that some of these stars appear in observations to experience a complete loss of mode energy in the core regions.
Although we cannot claim with certainty that red giant stars with low multipole mode visibility exhibit finite damping rates in their cores, the theoretical framework presented here provides a coherent and simple explanation for the presence or absence of the mixed mode signature in subsamples of these stars. In-depth theoretical studies are needed to assess whether damping mechanisms that inherently exhibit these properties exist or whether additional effects need to be taken into account.
Steps in this direction have been taken based on the assumption that a strong internal magnetic field dampens the oscillations, for example by \citet{fuller+15}, \citet{lecoanet+17}, \citet{rui+fuller23}, and \citet{david+25}.
Following on from the present work, we will next investigate the visibility and detectability of the mixed mode and multiplet signature along the evolution of red giant stars, testing the damping rates corresponding to the various damping mechanisms proposed in the literature \citep[e.g.,][]{dziembowski+01, dupret+09,fuller+15,muller+25}.

\begin{acknowledgements}
We thank the anonymous referee for their comments, which helped to improve the article.
We acknowledge funding from the ERC Consolidator Grant DipolarSound (grant agreement \# 101000296). In addition, we acknowledge support from the Klaus Tschira Foundation. 
\end{acknowledgements}

\bibliographystyle{aa}
\bibliography{ref}

\begin{appendix}

\section{Parameters describing mode damping} \label{app: Parameters describing mode damping}

In this work, as well as in the literature in general, various parameters are used to describe the efficiency of the damping processes that influence the oscillations of stars. In addition, the treatment of damping processes is complicated by the fact that some relationships between these parameters only apply under certain assumptions for the system containing the waves. For the reader's convenience, we summarize the interrelations between these parameters and the underlying assumptions in this section.

\subsection{One cavity}

In an arrangement with a single cavity in which the waves are damped at either the upper or lower boundary, but not at both boundaries, the following relationship applies between the parameters describing the damping \citep[e.g.,][and Eqs. \eqref{eq: amplitude modification factor}, \eqref{eq: resonance condition 1 cavity imag}, and \eqref{eq: amplitdue modification vs. eta}]{hekker+17}:
\begin{gather}
    \eta = -2\pi\nu_\Im = \frac{\mu}{t_{\rm cross}} = -\frac{\ln(|R|)}{2t_{\rm cross}} = \pi\Gamma = \tau^{-1}.
    \label{eq: damping parameters 1 cavitiy 1 damping}
\end{gather}
In this expression, we have used several parameters already defined in this paper, such as the damping rate $\eta$, the imaginary part of the cyclic eigenfrequency $\nu_\Im$, the amplitude modification factor $\mu$, the crossing time of the cavity $t_{\rm cross}$, the reflection coefficient $R$, and the mode linewidth $\Gamma$.
We have also included the mode lifetime $\tau$, which is defined here as the e-folding time of the wave amplitude.

Next, we consider a system with a cavity in which energy losses occur at both the upper and lower boundaries (indices $_{\rm t}$ and $_{\rm b}$, see Sect. \ref{sect: one cavity system}). Equation \eqref{eq: resonance condition 1 cavity imag} shows that the damping rate of the wave now depends on both damping processes. The other parameters describing the wave can be calculated from the interrelations between the parameters, so that the equivalent to Eq. \eqref{eq: damping parameters 1 cavitiy 1 damping} is given by
\begin{gather}
    \eta = -2\pi\nu_\Im = \frac{\mu_{\rm b}+\mu_{\rm t}}{t_{\rm cross}} = -\frac{\ln(|R_{\rm b}|)+\ln(|R_{\rm t}|)}{2t_{\rm cross}} = \pi\Gamma = \tau^{-1}.
\end{gather}
Apart from these parameters, which describe the damping of the oscillation globally, we can use Eq. \eqref{eq: amplitdue modification vs. eta} to assign a local damping rate to each of the two damping processes individually:
\begin{gather}
    \eta_{\rm t} = \frac{\mu_{\rm t}}{t_{\rm cross}} = -\frac{\ln(|R_{\rm t}|)}{2t_{\rm cross}}, \\
    \eta_{\rm b} = \frac{\mu_{\rm b}}{t_{\rm cross}} = -\frac{\ln(|R_{\rm b}|)}{2t_{\rm cross}}.
\end{gather}
The global damping rate of the wave is given by the sum of the local damping rates corresponding to the individual damping processes (i.e., $\eta = \eta_{\rm t} + \eta_{\rm b}$). In this work, we use the imaginary part of the frequency $\nu_\Im$, the linewidth $\Gamma$, and the mode lifetime $\tau$ exclusively to denote the global damping of the wave.

\subsection{Two cavities}

In a two-cavity arrangement, where energy losses can occur at both the upper and lower boundaries (subscripts $_{\rm p}$ and $_{\rm g}$, see Sect. \ref{sect: two cavity system}), estimating the global parameters that describe the damping of the wave is more complex than in a single-cavity system. This is because the resonance condition (Eq. \eqref{eq: resonance condition 2 cavities}) cannot be split into a single equation describing the real and imaginary parts of the frequency. Nevertheless, Eq. \eqref{eq: resonance condition 2 cavities} clearly shows that the global parameters describing the damping  of the wave are influenced by both damping processes.

The local damping rates can be defined in a manner similar to that for the single-cavity system:
\begin{gather}
    \eta_{\rm p} = \frac{\mu_{\rm p}}{t_{\rm p,cross}} = -\frac{\ln(|R_{\rm p}|)}{2t_{\rm p,cross}}, \\
    \eta_{\rm g} = \frac{\mu_{\rm g}}{t_{\rm g,cross}} = -\frac{\ln(|R_{\rm g}|)}{2t_{\rm g,cross}}.
\end{gather}
The only conceptual difference to the single-cavity system is that we now have to take into account the different crossing times of the two cavities. Crucially, however, the global damping rate of the wave $\eta$ cannot be calculated as the sum of the local damping rates $\eta_{\rm p}$ and $\eta_{\rm g}$ in the two-cavity system. This is because $\eta$ also depends on the sensitivity of the oscillations to the two cavities, which differs from one mixed mode to another. 

In the limiting case without damping (i.e., $|R_{\rm p}| = |R_{\rm g}| = 1$), the sensitivity of a given mixed mode to the g-mode cavity can be estimated as a parameter $\zeta$ \citep[Eq. \eqref{eq: zeta},][]{goupil+13,deheuvels+15}. Using this parameter, the global damping rate of the wave can be approximated as \citep[e.g.,][]{grosjean+14,hekker+17}
\begin{gather}
    \eta \approx (1-\zeta)\ \eta_{\rm p} + \zeta\ \eta_{\rm g}.
    \label{eq: damping rate zeta}
\end{gather}
In fact, this expression can be derived from Eq. \eqref{eq: resonance condition 2 cavities} using a first-order perturbation approach, where it is assumed that the energy loss due to the damping processes is small \citep[i.e., $\mu_{\rm p}, \mu_{\rm g} \ll 1$, see section 4.1 of][]{takata16b}.
However, Eq. \eqref{eq: resonance condition 2 cavities} should be used instead of Eq. \eqref{eq: damping rate zeta} if the assumption that the efficiency of the damping processes is low is not justified.

\section{Power spectral density of waves with real frequencies} \label{app: Power spectral density}

\subsection{Derivation}

The PSD $P(\omega)$ of a temporal signal $S(t)$ is given by
\begin{gather}
    P(\omega) = \lim_{\mathcal{T}\rightarrow \infty} \frac{1}{\mathcal{T}} \left|\Hat{S}(\omega)\right|^2 \propto \lim_{\mathcal{T}\rightarrow \infty} \frac{1}{\mathcal{T}} \left|\int_{-\mathcal{T}/2}^{\mathcal{T}/2} S(t) e^{\mathrm{i}\omega t} \ {\rm d}t \right|^2.
\end{gather}
Here, we set the signal to the wave function in Eq. \eqref{eq: wave function arrows} evaluated at the photometric radius $R_{\rm phot}$, which is closely related to the radial component of the observable displacement field. In the p-mode cavity, it can be expressed as follows:
\begin{gather}
    \Psi_{\omega_0}(R_{\rm phot},t) = \frac{1}{\sqrt{k_{r,\omega_0}}} \left( a_{\omega_0}^\rightarrow e^{\mathrm{i} \varphi_{\omega_0}}+ a_{\omega_0}^\leftarrow e^{-\mathrm{i} \varphi_{\omega_0}}\right) e^{-\mathrm{i}{\omega_0} t}.
\end{gather}
We assume that the frequency of the wave $\omega_0$ is real and the subscript $_{\omega_0}$ indicates dependence on the frequency of the wave. In the following, we denote $\Psi_{\omega_0}(t) \equiv \Psi_{\omega_0}(R_{\rm phot},t)$.
As a first step, we calculate the Fourier transform of the exponential function that contains the dependence on time:
\begin{gather}
    \int_{-\mathcal{T}/2}^{\mathcal{T}/2} e^{-\mathrm{i}(\omega_0 - \omega) t}\ {\rm d}t \propto \frac{\sin\left(\frac{\mathcal{T}}{2}(\omega-\omega_0)\right)}{\omega-\omega_0}.
\end{gather}
The squared modulus of the Fourier-transformed wave function is then given by
\begin{gather}
    \left|\Hat{\Psi}_{\omega_0}(\omega)\right|^2 \propto \frac{1}{k_{r,\omega_0}}\frac{\sin^2\left(\frac{\mathcal{T}}{2}(\omega-\omega_0)\right)}{(\omega-\omega_0)^2}  \notag\\
    \qquad \cdot \left[ |a_{\omega_0}^\rightarrow|^2 + |a_{\omega_0}^\leftarrow|^2 + 2\left( a_{\omega_0}^\rightarrow(a_{\omega_0}^\leftarrow)^\star e^{\mathrm{i} 2 \varphi_{\omega_0}} \right)_\Re \right].
\end{gather}
After multiplying by $1/\mathcal{T}$ and taking the limit as $\mathcal{T} \rightarrow \infty$, the PSD of a wave with frequency $\omega_0$ reads
\begin{gather}
    P_{\omega_0}(\omega) \propto \frac{\delta(\omega-\omega_{0})}{k_{r,\omega_0}}  \left[ |a_{\omega_0}^\rightarrow|^2 + |a_{\omega_0}^\leftarrow|^2 +\ 2\left( a_{\omega_0}^\rightarrow(a_{\omega_0}^\leftarrow)^\star e^{\mathrm{i} 2 \varphi_{\omega_0}} \right)_\Re \right],
\end{gather}
where $\delta(\omega-\omega_{0})$ denotes the Dirac delta distribution localized at $\omega = \omega_0$. In this step, we used the following identity \citep[e.g., equation (5.290) of][]{sakurai+napolitano20}:
\begin{gather}
     \lim_{\mathcal{T}\rightarrow \infty} \frac{\sin^2\left(\frac{\mathcal{T}}{2}(\omega-\omega_0)\right)}{\mathcal{T}(\omega-\omega_0)^2} \propto \delta(\omega-\omega_{0}).
\end{gather}
To obtain the complete PSD, which contains the contribution of all waves with different frequencies, we integrate the PSD of a single wave over all real frequencies $\omega_0$, which yields
\begin{gather}
    P(\omega) \propto \frac{1}{k_{r,\omega}}  \left[ |a_{\omega}^\rightarrow|^2 + |a_{\omega}^\leftarrow|^2 + 2\left( a_{\omega}^\rightarrow(a_{\omega}^\leftarrow)^\star e^{\mathrm{i} 2 \varphi_{\omega}} \right)_\Re \right],
    \label{eq: complete PSD with cross term}
\end{gather}
where the subscript $_{\omega}$ indicates dependence on the frequency $\omega$.

Equation \eqref{eq: complete PSD with cross term} contains a cross term of the amplitudes. In the following subsection (Appendix \ref{app: impact of the cross term}), we show that this cross term is a source of asymmetry for the peaks in the PSD and leaves the general resonance pattern largely unchanged. In particular, neglecting the cross term has no significant effect on the predicted visibilities and brings the corresponding power spectrum closer to a sum of Lorentzian functions (see Figs. \ref{fig: IRE_Lorentz_1cavity} and \ref{fig: IRE_Lorentz_2cavities}), which is commonly used to calculate synthetic PSDs and, by design, only features symmetric peaks.
For these reasons, we neglect the cross term in the rest of this study, which means that our expression for the complete PSD is reduced to
\begin{gather}
    P(\omega) \propto C_\omega \left(|a_{\omega}^\rightarrow|^2 + |a_{\omega}^\leftarrow|^2 \right),
    \label{eq: PSD without cross term}
\end{gather}
where $C_\omega$ is a factor that depends on $\omega$.

\subsection{Impact of the cross term} \label{app: impact of the cross term}

\begin{figure}[]
    
    \resizebox{\hsize}{!}{\includegraphics{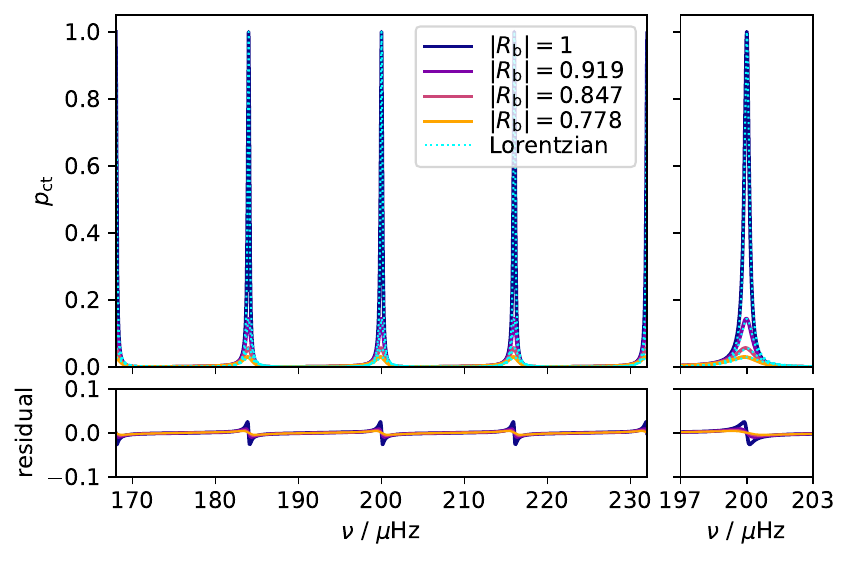}}
    \caption{Same as Fig. \ref{fig: IRE_Lorentz_1cavity}, now including the cross term in Eq. \eqref{eq: normalized PSD cross term 1 cavity}.}
    \label{fig: IRE_Lorentz_1cavity_crossterm}
\end{figure}
\begin{figure}[]
    
    \resizebox{\hsize}{!}{\includegraphics{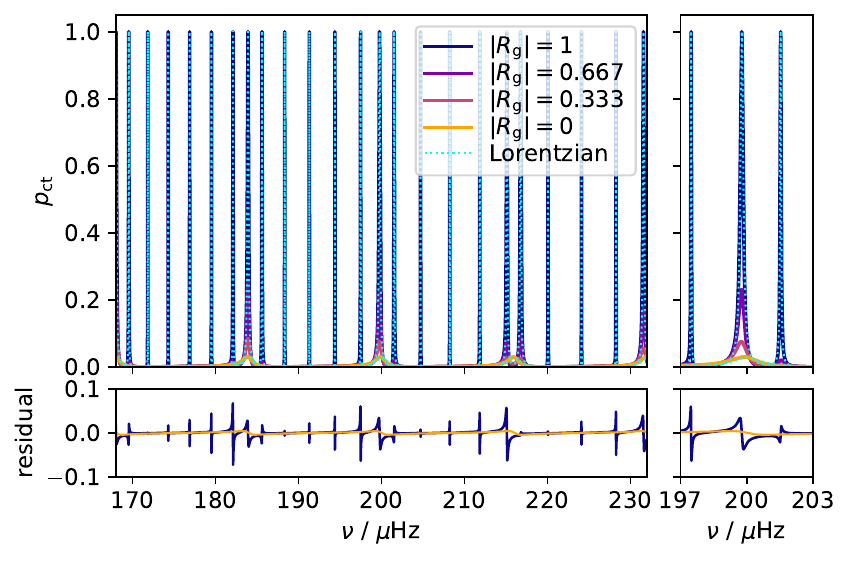}}
    \caption{Same as Fig. \ref{fig: IRE_Lorentz_2cavities}, now including the cross term in Eq. \eqref{eq: normalized PSD cross term 2 cavities}.}
    \label{fig: IRE_Lorentz_2cavities_crossterm}
\end{figure}

In this section, we briefly discuss the influence of the cross term of the amplitudes derived in the previous subsection.
If we express the amplitudes of the upward- and downward-traveling waves as $a_{\omega}^\rightarrow = |a_{\omega}^\rightarrow| e^{\mathrm{i}\alpha_{\omega}^\rightarrow}$ and $a_{\omega}^\leftarrow = |a_{\omega}^\leftarrow| e^{\mathrm{i} \alpha_{\omega}^\leftarrow}$, we can rewrite the complete PSD as
\begin{gather}
    P(\omega) \propto \frac{1}{k_{r,\omega}}  \left[ |a_{\omega}^\rightarrow|^2 + |a_{\omega}^\leftarrow|^2 \right. \notag\\
    \qquad \left. +\ 2\ |a_{\omega}^\rightarrow||a_{\omega}^\leftarrow|\cos\left(\alpha_{\omega}^{\rightarrow} - \alpha_{\omega}^\leftarrow + 2 \varphi_\omega \right) \right].
\end{gather}
This means that the cross term depends on the relationship between the phases of the upward- and downward-traveling waves and on $\varphi_\omega$.

In Eq. \eqref{eq: normalized power spectrum 2 cavities}, we define the normalized power spectrum $p$ corresponding to a configuration with one and two cavities using the expression of the PSD as defined in Eq. \eqref{eq: PSD without cross term} (i.e., without the cross term). If we include the cross term, the normalized power spectrum instead becomes
\begin{gather}
   p_{\rm ct} = \frac{A_1}{\mathcal{H}_{\rm t}} \left[1 + |R_{\rm b}|^2  -\ 2 |R_{\rm b}| \cos\left(- \delta_{\rm b} + 2 \varphi_{\omega,-} \right)\right],
\end{gather}
in the single-cavity system and using Eq. \eqref{eq: ampl up down relation}
\begin{gather}
   p_{\rm ct} = \frac{A_2}{\mathcal{H}_{\rm p}} \left[1 + e^{-4(\Phi_{\rm g})_\Im} \right.\notag\\
   \qquad \left. -\ 2\ e^{-2(\Phi_{\rm g})_\Im} \cos\left(- \delta_q + 2(\Phi_{\rm g})_\Re + 2 \varphi_{\omega,-} \right)\right],
\end{gather}
in the two-cavity system. The subscript $_{\rm ct}$ stands for  "cross term" and the subscript $_-$ of $\varphi_{\omega,-}$ indicates that we have chosen the reference radius of the phase as the lower boundary of the p-mode cavity (i.e., $\Bar{r} =r_{\rm -}$ or $r_{\rm p,-}$ in the single- and two-cavity configuration, respectively).
Next, we assume that the photometric radius is close to the upper boundary of the p-mode cavity, which means that $R_{\rm phot} \approx r_{\rm +}$ or $r_{\rm p,+}$. Because of this, the phase $\varphi_{\omega,-}$ can be identified with $\Theta$ or $\Theta_{\rm p}$ (see Eq. \ref{eq: theta}). Furthermore, we set the phase shifts to $\delta_{\rm b} \approx \pi/2$ and $\delta_q \approx \pi/2$, where we have used the weak coupling approximation for the latter \citep[see][]{pincon+takata22}, so that the normalized power spectrum including the phase term can be written as
\begin{gather}
   p_{\rm ct} \approx \left[1 + |R_{\rm b}|^2  - 2\ |R_{\rm b}| \sin\left(2 \Theta \right)\right] \frac{A_1}{\mathcal{H}_{\rm t}},
   \label{eq: normalized PSD cross term 1 cavity}
\end{gather}
in the single-cavity system and
\begin{gather}
   p_{\rm ct} \approx \left[1 + e^{-4(\Phi_{\rm g})_\Im} - 2\ e^{-2(\Phi_{\rm g})_\Im} \sin\left( 2\left[ (\Phi_{\rm g})_\Re + \Theta_{\rm p} \right]\right)\right] \frac{A_2}{\mathcal{H}_{\rm p}}
   \label{eq: normalized PSD cross term 2 cavities}
\end{gather}
in the two-cavity system.

The resulting power spectrum $p_{\rm ct}$ can be explicitly calculated for a given set of input values. In Figs. \ref{fig: IRE_Lorentz_1cavity_crossterm} and \ref {fig: IRE_Lorentz_2cavities_crossterm}, we show $p_{\rm ct}$ for the same parameters as the normalized power spectra shown in Figs. \ref{fig: IRE_Lorentz_1cavity} and \ref{fig: IRE_Lorentz_2cavities}. Figures \ref{fig: IRE_Lorentz_1cavity_crossterm} and \ref{fig: IRE_Lorentz_2cavities_crossterm} demonstrate that the inclusion of the cross term does not strongly affect the general resonance pattern and acts as a source of asymmetry for the peaks (see Appendix \ref {app: Asymmetric peaks}). Therefore, neglecting the cross term has no significant impact on the analysis conducted in this study.

\section{Asymmetric peaks} \label{app: Asymmetric peaks}

Solar p-modes are known to be asymmetrical \citep[e.g.,][]{nigam+kosovichev98,chaplin+appourchaux99}. There are thought to be two reasons for this asymmetry \citep{basu&chaplin17}.
The first contribution to this asymmetry arises from the excitation mechanism of the modes. On one hand, it is because the energy injected into the individual modes is not independent of that supplied to the other modes. On the other hand, the asymmetry is the result of the correlation of the excitation with the background noise in the PSD. However, the mode excitation is not correlated in our simplified theoretical framework, which means that this cannot be the source of the asymmetry of the mixed modes computed here.

The second contribution is the interference of waves that traveled through the star along different paths before interfering. The difference in path results in a phase shift between the waves, which is responsible for the asymmetry.
To examine the second contribution within a simple mathematical model, \citet{basu&chaplin17} propose the following scenario. Consider an arrangement with one cavity as presented in Sect. \ref{sect: One and two cavity systems}, where the excitation does not occur at the upper boundary $r_{\rm p,+}$ but at a different radius $r_{\rm exci}$ in the cavity (i.e., $r_{\rm p,-}< r_{\rm exci} < r_{\rm p,+}$). Instead of only a downward-propagating wave caused by the excitation, we now assume that we have an upward- and downward-traveling wave with the same amplitude. The upward-traveling wave is eventually reflected at the upper boundary of the cavity and itself becomes a downward-traveling wave. By traveling from $r_{\rm exci}$ to $r_{\rm p,+}$ and back, it has then acquired the phase shift $2\varphi_{\rm exci}$ compared to the wave that originally traveled downward, where $\varphi_{\rm exci}$ is given by
\begin{gather}
    \varphi_{\rm exci} \equiv \varphi(r_{\rm exci}; r_{\rm p,+}) = \int_{r_{\rm p,+}}^{r_{\rm exci}} k_{r}\ {\rm d}r.
\end{gather}
To simplify this expression, we assume that the cavity is a p-mode cavity and make the assumption $\omega^2 \gg N^2, S_\ell^2$. The local radial wave number can then be approximated as
\begin{gather}
    k_{r} \approx \frac{\omega}{c_{\rm s}},
\end{gather}
such that the phase shift can be written as
\begin{gather}
    \varphi_{\rm exci} \approx \omega \int_{r_{\rm p,+}}^{r_{\rm exci}}\frac{1}{c_{\rm s}}\ {\rm d}r \equiv \omega\ \frac{\Delta t}{2},
\end{gather}
where we introduced the separation in time of the two downward-traveling waves $\Delta t$. The internal resonance enhancement factor of a system with such a phase shift (subscript $_{\rm ps}$) is given by
\begin{gather}
    A_{\rm 1,ps} = \frac{|a^\leftarrow_+ + a^\leftarrow_{\rm +,ps}|^2}{2|a_{\rm exci}|^2} = \frac{1}{2}\left|1 + e^{\mathrm{i}2(\frac{\delta_{\rm p}}{2} + \omega\frac{\Delta t}{2} + \mathrm{i}\mu_{\rm p})}\right|^2 A_1.
\end{gather}
The scaling factor in this expression is equivalent to that in equation (5.53) of \citet{basu&chaplin17} if the phase shift and the energy loss due to the reflection at the upper boundary are neglected (i.e., $\delta_{\rm p} = \mu_{\rm p} = 0$). This can be seen by expressing the squared absolute value of the exponential function as a cosine function, such that
\begin{gather}
    A_{\rm 1,ps} = \left(1 + \cos(\omega\Delta t)\right)\ A_1,
    \label{eq: asymmetric IRE}
\end{gather}
where we assumed that $\omega$ is real, which is in line with our discussion in Sect. \ref{sect: Synthetic power spectrum}. The factor with the cosine function indeed induces asymmetries on the peaks of $A_1$ when $\Delta t$ is chosen appropriately. This indicates that the asymmetries of the mixed modes might also be the result of a frequency-dependent phase shift between two waves that interfere with each other.

In the case of mixed modes in the two-cavity configuration, this phase shift is not induced during the excitation of the waves, but rather during the reflection at the lower boundary of the p-mode cavity. Figure \ref{fig: sketch 2cavities} shows that the part of the downward-traveling wave, which is reflected at the lower boundary (i.e., $a_{\rm p,ref}$), interferes with the part that tunnels back from the g- into the p-mode cavity (i.e., $a_{\rm p,coup}$). In the g-mode cavity, the wave undergoes an infinite number of reflections, while a fraction of its energy leaks back into the p-mode cavity with each round trip. For this reason, $a_{\rm p,coup}$ has accumulated a phase shift relative to $a_{\rm p,ref}$, and we obtain a scenario comparable to that discussed for the p-modes.
However, the partial phase shift is now introduced at the lower boundary of the p-mode cavity. This means that we cannot express the influence of the phase shift as an additional factor to the internal resonance enhancement factor $A_1$, as in Eq. \eqref{eq: asymmetric IRE}, but must replace $A_1$ with a function that takes the phase shift into account internally (such as $A_2$).

We conclude that the asymmetric peaks of the mixed modes are caused by the phase shift accumulated by the part of the wave transmitted through the evanescent zone and back, compared to the part directly reflected at it. It is therefore an important feature of the oscillations, which is also expected to occur in the PSD of real stars, although it is probably not resolvable. In that sense, the approach to estimating the power spectrum presented in Sect. \ref{sect: Synthetic power spectrum} contains more information than a sum of symmetric Lorentz functions, which explains the skewness of the peaks in Fig. \ref{fig: IRE_Lorentz_2cavities}.

Note that in Appendix \ref{app: Power spectral density}, we discuss another cause of asymmetry of the peaks in the power spectrum, which is neglected in the rest of this study. This other type of asymmetry even occurs in the power spectrum of the single-cavity configuration (see Fig. \ref {fig: IRE_Lorentz_1cavity_crossterm}) and can be expressed as an additional factor to the internal resonance enhancement factor, similar to Eq. \eqref{eq: asymmetric IRE} (see Eqs. \eqref{eq: normalized PSD cross term 1 cavity} and \eqref{eq: normalized PSD cross term 2 cavities}).

\section{Variation in the visibility without core damping} \label{app: Variation in visibility no core damping}

\begin{figure}[h]
    
    \resizebox{\hsize}{!}{\includegraphics{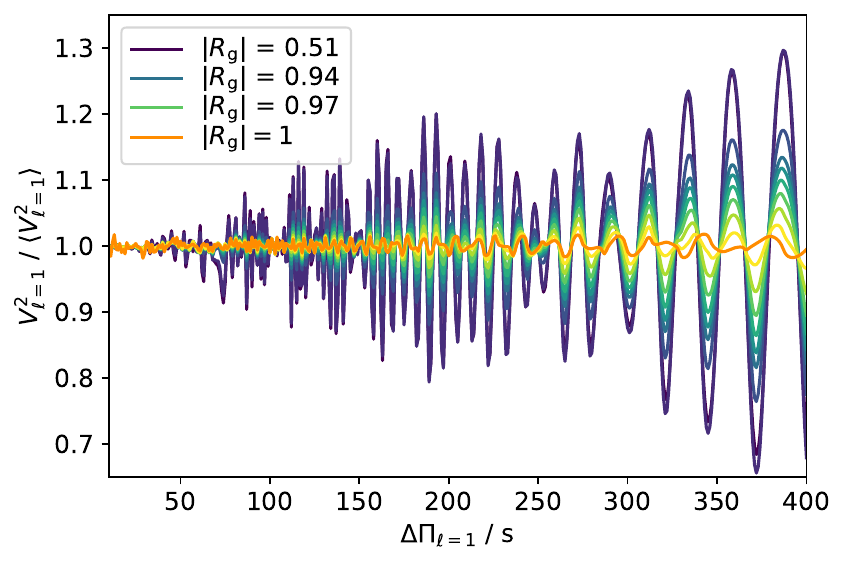}}
    \caption{Relative variation of the dipole mode visibility as a function of period spacing for the set of stellar parameters A with $|R_{\rm p}| = 0.95$. Colors indicate different values of $|R_{\rm g}|$ ranging from 0.51 to 1. The variable $\langle V_\ell^2\rangle$ denotes the average of the visibility over all tested values of $\Delta\Pi_{\ell=1}$ for each value of $|R_{\rm g}|$.}
    \label{fig: relative visibility umdamped}
\end{figure}

\citet{mosser+17} argue that the total visibility of the mixed modes of a given spherical degree $\ell$ in one pressure radial order should be equal to one if no damping process operates in the g-mode cavity (i.e., if $|R_{\rm g}|=1$). 
However, when calculating the visibility according to the method presented in Sect. \ref{sect: Synthetic power spectrum}, we find that although the visibility is indeed close to one, the exact value depends on the stellar parameters and varies by a few percent. An example of the variation of the visibility with the period spacing $\Delta\Pi_{\ell=1}$ is shown in Fig. \ref{fig: relative visibility umdamped}.
While the variation in the visibility corresponds to how well the nominal p- and g-mode frequencies align in the observable frequency range when $|R_{\rm g}|<1$ (see Sect. \ref{sect: results parameter study} and Appendix \ref {app: Additional figures}), this does not appear to be the decisive factor when $|R_{\rm g}|=1$. In particular, for larger values of $\Delta\Pi_{\ell=1}$, Fig. \ref{fig: relative visibility umdamped} clearly shows that the qualitative changes in the visibility with $\Delta\Pi_{\ell=1}$ when $|R_{\rm g}|=1$ do not align with those that can be observed for $|R_{\rm g}|<1$.
To understand why this is the case, we examine the extent to which the assumptions made by \citet{mosser+17} are fulfilled in our computations.

\subsection{Underlying assumption}

If no energy loss occurs in the g-mode cavity, the visibility can be expressed as \citep{dupret+09,benomar+14,grosjean+14,mosser+17}
\begin{gather}
    V^2_\ell = \sum_{n_{\rm mix}} (1-\zeta),
    \label{eq: visibility zeta}
\end{gather}
where $n_{\rm mix}$ is the number of mixed modes of the spherical degree $\ell$ in one pressure radial order and $\sum_{n_{\rm mix}}$ denotes the sum over these mixed modes.
The parameter $\zeta$ is a function describing the sensitivity of a given mixed mode to the properties of the g-mode cavity \citep{goupil+13,deheuvels+15}.
The main assumption made by \citet{mosser+17} is that the sum of $\zeta$ over all mixed modes in one pressure radial order is equal to the number of nominal g-modes in the same radial order. This can be expressed as \citep[see also][]{mosser+15}
\begin{gather}
     \sum_{n_{\rm mix}} \zeta \approx n_{\rm mix} -1.
\end{gather}
To verify the validity of this assumption in our calculations, we rewrite it into a parameter $v$ that quantifies the extent to which it is violated:
\begin{gather}
      v \equiv n_{\rm mix} - 1 - \sum_{n_{\rm mix}} \zeta \approx 0.
      \label{eq: assumption}
\end{gather}
If $v$ is equal to zero, the assumption is valid. If $v$ is not equal to zero, the assumption does strictly speaking not hold.

\subsection{Testing the assumption}

\begin{figure}[]
    
    \resizebox{\hsize}{!}{\includegraphics{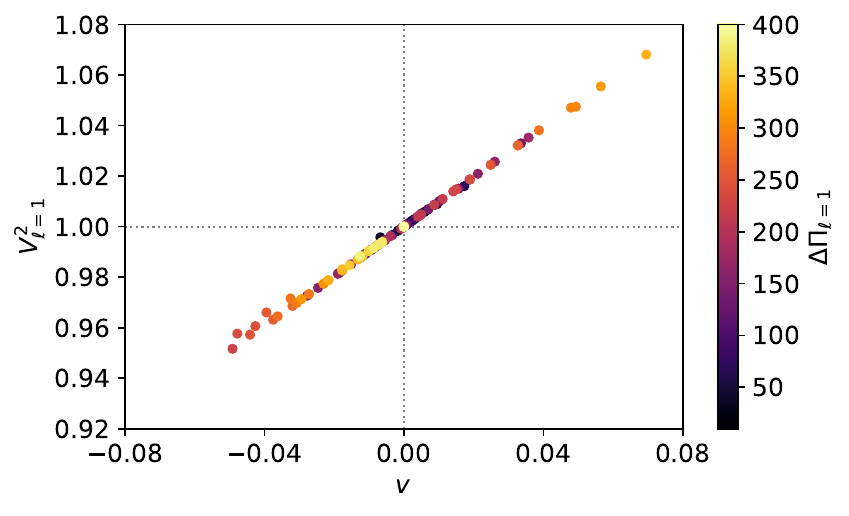}}
    \caption{Dipole mode visibility of the central pressure radial order (i.e., 192 to 208 $\mu$Hz) calculated using the normalized power spectrum $p$ for the set of stellar parameters A. The modulus of the reflection coefficients of the outer boundaries are set to $|R_{\rm p}|=0.95$ and $|R_{\rm g}|=1$. Colors indicate the period spacing. The values on the horizontal axis correspond to deviations from assumption made by \citet{mosser+17} measured by $v$ (Eq. \eqref{eq: assumption}).}
    \label{fig: assumptions mosser}
\end{figure}

In this section, we test the effect of the aforementioned assumption on the dipole mode visibility using the parameters from the set A in Table \ref{tab: sets of stellar parameters} in a scenario similar to that considered by \citet{mosser+17}. Therefore, we only consider one pressure radial order. We select the central one, which ranges from 192 to 208 $\mu$Hz and is bounded by two radial modes. The position of one of the nominal dipolar p-modes lies exactly in the middle of this frequency range, at 200 $\mu$Hz. In addition, we neglect the influence of the Gaussian function on the power spectrum described in Sect. \ref{sect: relative power spectrum}, which means that we essentially use the normalized power spectrum $p$ instead of the relative power spectrum $\mathcal{P}$ to calculate the visibility for this section only.

To evaluate Eq. \eqref{eq: assumption}, we must constrain $n_{\rm mix}$ and $\zeta$. The number of mixed modes $n_{\rm mix}$ can be determined by counting the local maxima of the normalized power spectrum $p$ for $\ell=1$ in the frequency range under consideration. 
The parameter $\zeta$ is analytically given by \citep{goupil+13,deheuvels+15}
\begin{gather}
    \zeta = \left( 1 + \frac{\nu^2\ \Delta\Pi_\ell}{\Delta\nu} \frac{\sin(2 \Theta_{\rm g})}{\sin(2 \Theta_{\rm p})}\right)^{-1}.
    \label{eq: zeta}
\end{gather}
Equation \eqref{eq: zeta} has been derived under the assumption that there is no damping whatsoever (i.e., $|R_{\rm p}| = |R_{\rm g}| = 1$). Both in our computations and in reality this is never truly the case because the waves are always damped at the upper region of the p-mode cavity.
However, Eq. \eqref{eq: zeta} still holds approximately when the energy losses of the wave are small, which is why we use it to evaluate $\zeta$ for each mixed mode in the frequency range under consideration.
To estimate the phases $\Theta_{\rm p}$ and $\Theta_{\rm g}$, we use Eqs. \eqref{eq: theta_p with ell} and \eqref{eq: frequency shift theta_g} with $\delta\nu=0$. The frequencies $\nu$ of the mixed modes can be determined as the frequencies of the local maxima of $p$.

In Fig. \ref{fig: assumptions mosser}, we show that the variation of the dipole mode visibility changes depending on how well the assumption made by \citet{mosser+17} is satisfied. We can see clearly that when the assumption is true, which means that $v$ is equal to zero, the visibility is consistent with one. The deviations of the visibility from unity do indeed correlate with larger deviations of $v$ from zero, demonstrating that the variation in the visibility without damping in the g-mode cavity is caused by the fact that the simplifying assumption of \citet{mosser+17} is not always truly valid.
The behavior of the visibility shown in Fig. \ref{fig: assumptions mosser} is representative for different pressure radial orders and other choices for the stellar parameters.

Considering the results shown in Fig. \ref{fig: assumptions mosser}, it seems likely that the deviations of the visibility from one in the case that $|R_{\rm g}|=1$ shown in Figs. \ref{fig: visibility versus R_g} and \ref{fig: relative visibility umdamped} are also mainly caused by violations of this assumption. However, three additional aspects must be taken into account in the calculations presented in this article, which we will not discuss in detail here as they would go beyond the scope of this work. First, the frequency range considered in this work spans six pressure radial orders instead of just one. Second, the boundaries of the frequency range do not always coincide with the location of the two radial modes, as this depends on the individual values of $\nu_{\rm max}$ and $\Delta\nu$. Third, the Gaussian envelope influences the amplitudes of both the mixed and radial modes and therefore also contributes to the variation of the visibility. 
It appears that these three factors slightly lower the fluctuations in the visibility when $|R_{\rm g}|=1$. While the visibility varies by up to 6\% in Fig. \ref{fig: assumptions mosser}, the deviation from unity is much smaller in Fig. \ref{fig: relative visibility umdamped}. This suggests that the assumption that the visibility is equal to one when there is no damping in the g-mode cavity is most likely justified for practical purposes.

\section{Parameter study} \label{app: Parameter study}

\begin{figure*}[h]
    
    \resizebox{\hsize}{!}{\includegraphics{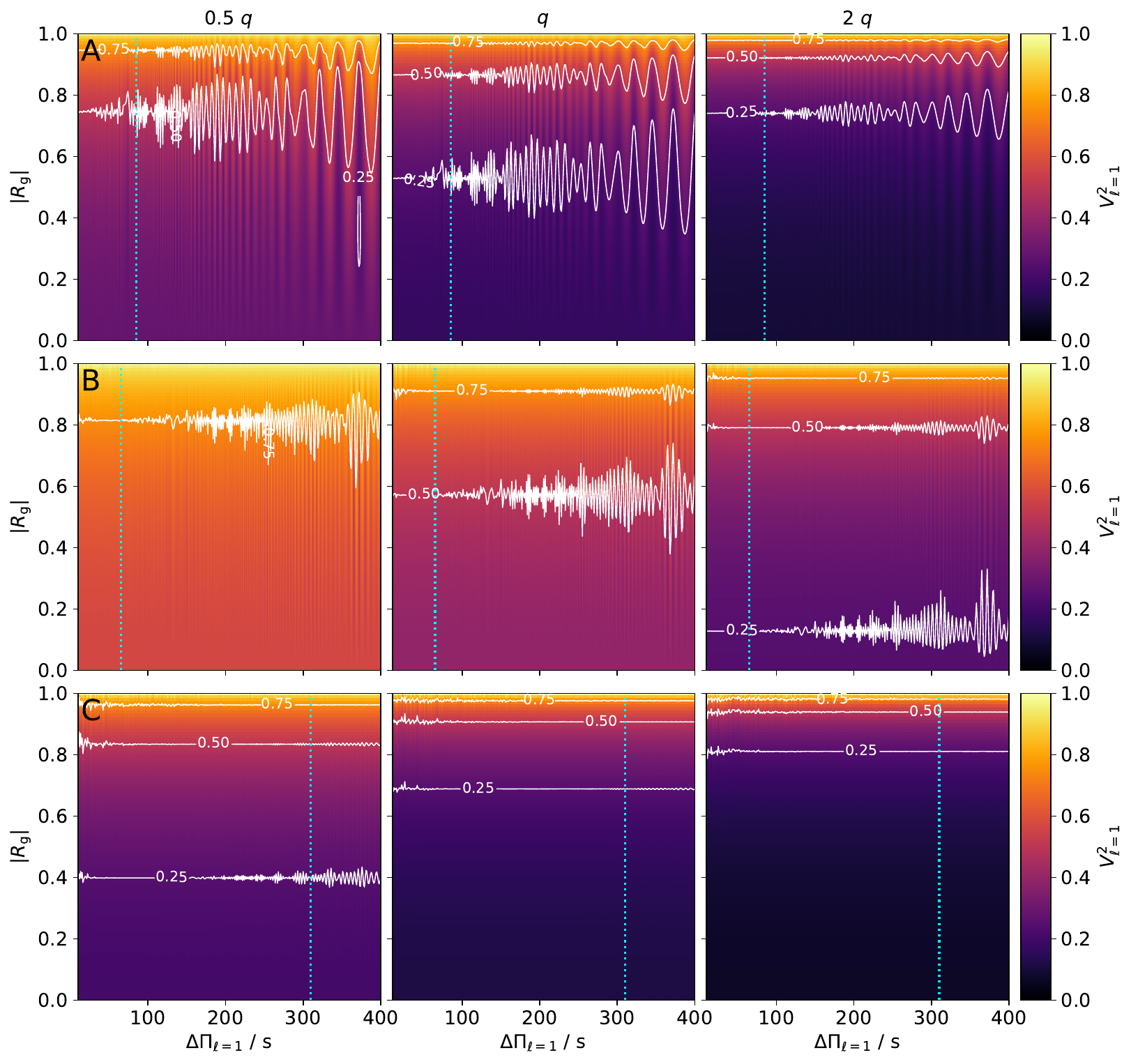}}
    \caption{Same as Fig. \ref{fig: 3x3 visibility}, now the columns correspond to different values in the coupling strength. The middle column is identical to the one in Fig. \ref{fig: 3x3 visibility}.}
    \label{fig: 3x3 visibility coupling}
\end{figure*}
\begin{figure*}[h]
    
    \resizebox{\hsize}{!}{\includegraphics{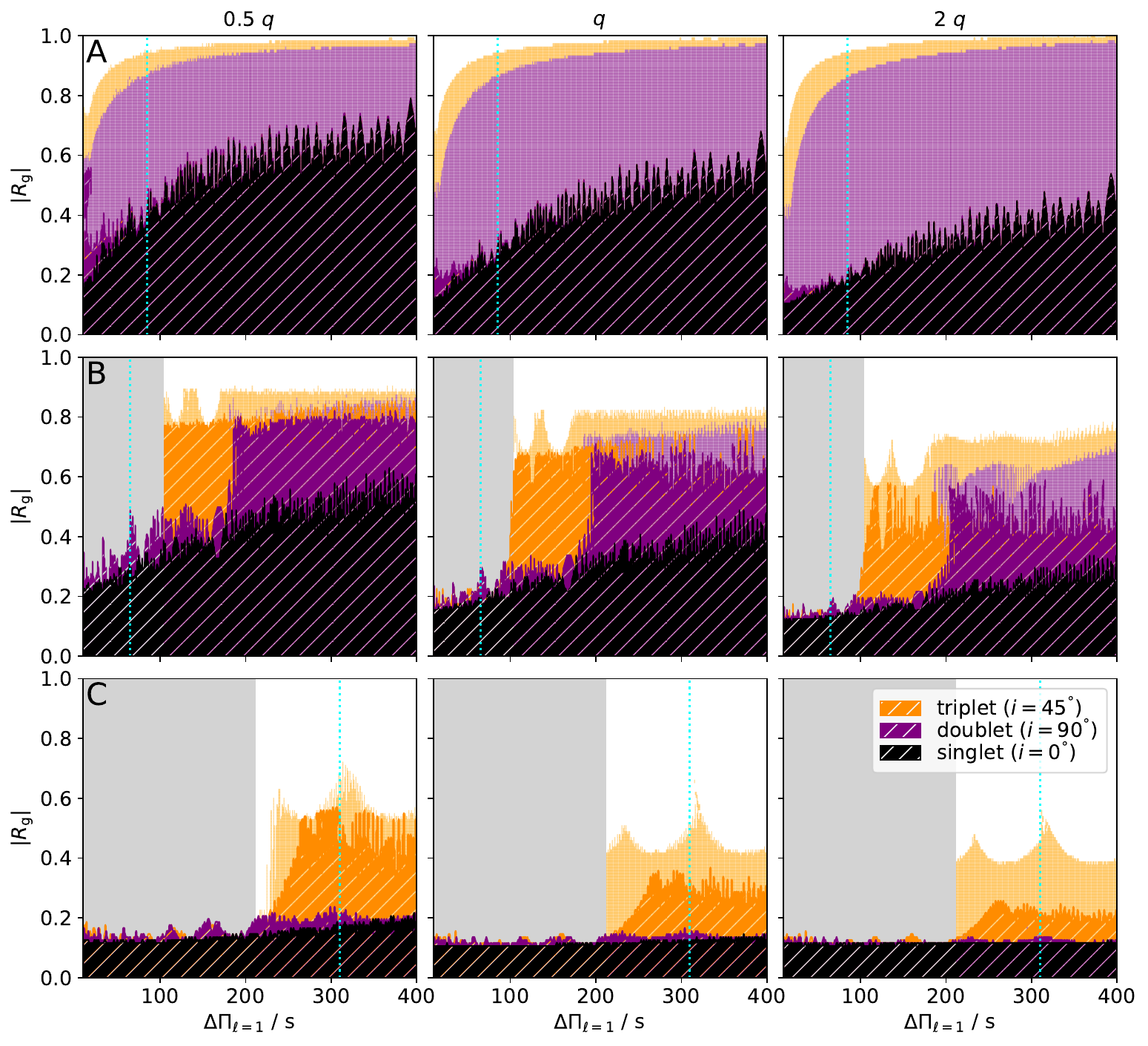}}
    \caption{Same as Fig. \ref{fig: 3x3 mixed mode detectability}, now the columns correspond to different values in the coupling strength. The middle column is identical to the one in Fig. \ref{fig: 3x3 mixed mode detectability}.}
    \label{fig: 3x3 mixed mode detectability coupling}
\end{figure*}

\begin{figure*}[h]
    
    \resizebox{\hsize}{!}{\includegraphics{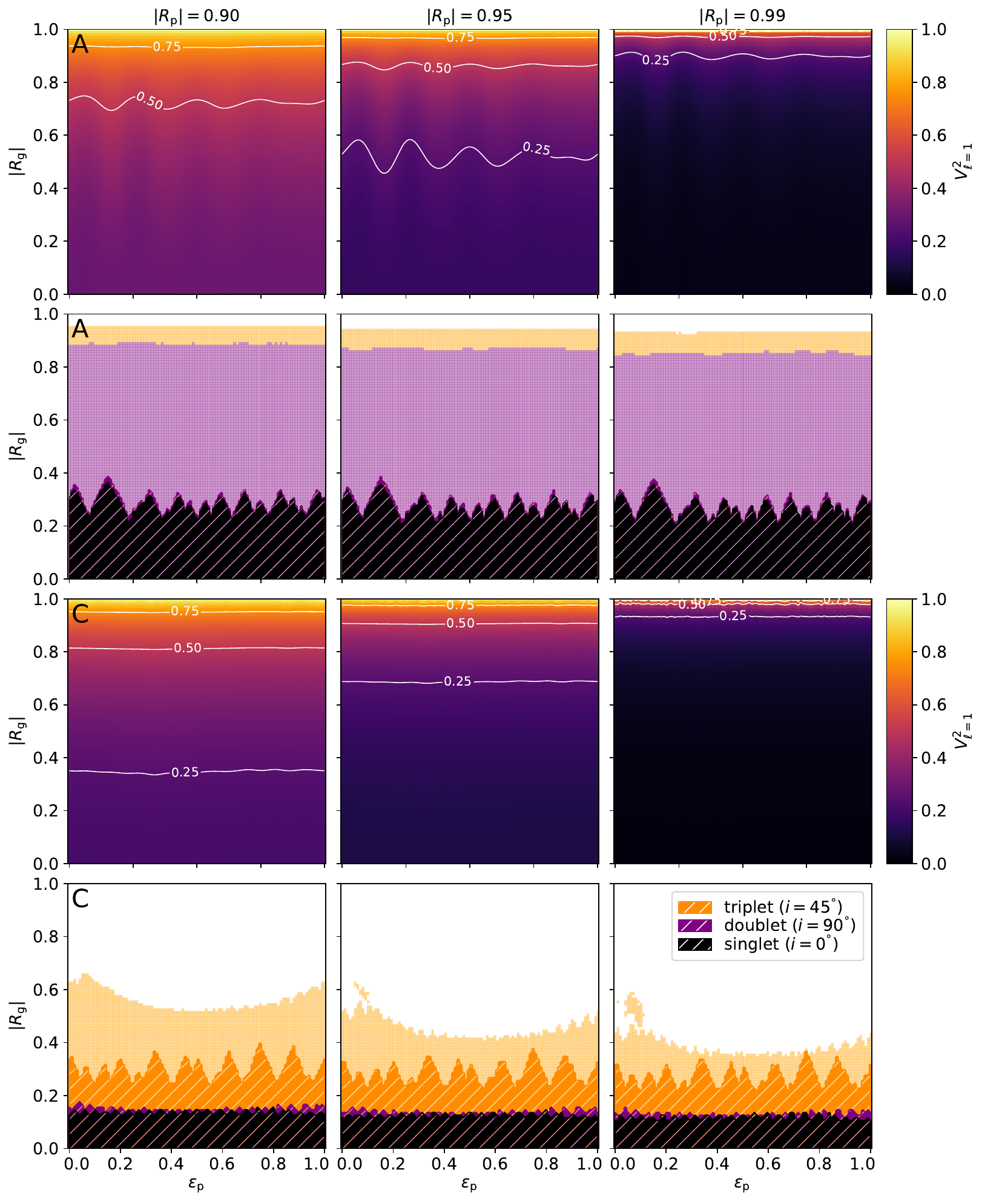}}
    \caption{Same as Figs. \ref{fig: 3x3 visibility} and \ref{fig: 3x3 mixed mode detectability}, now as a function of the phase shift in the p-mode cavity instead of the period spacing. The first two rows correspond to the set of stellar parameters A, and the last two rows correspond to set C.}
    \label{fig: 3x3 epsilon p}
\end{figure*}
\begin{figure*}[h]
    
    \resizebox{\hsize}{!}{\includegraphics{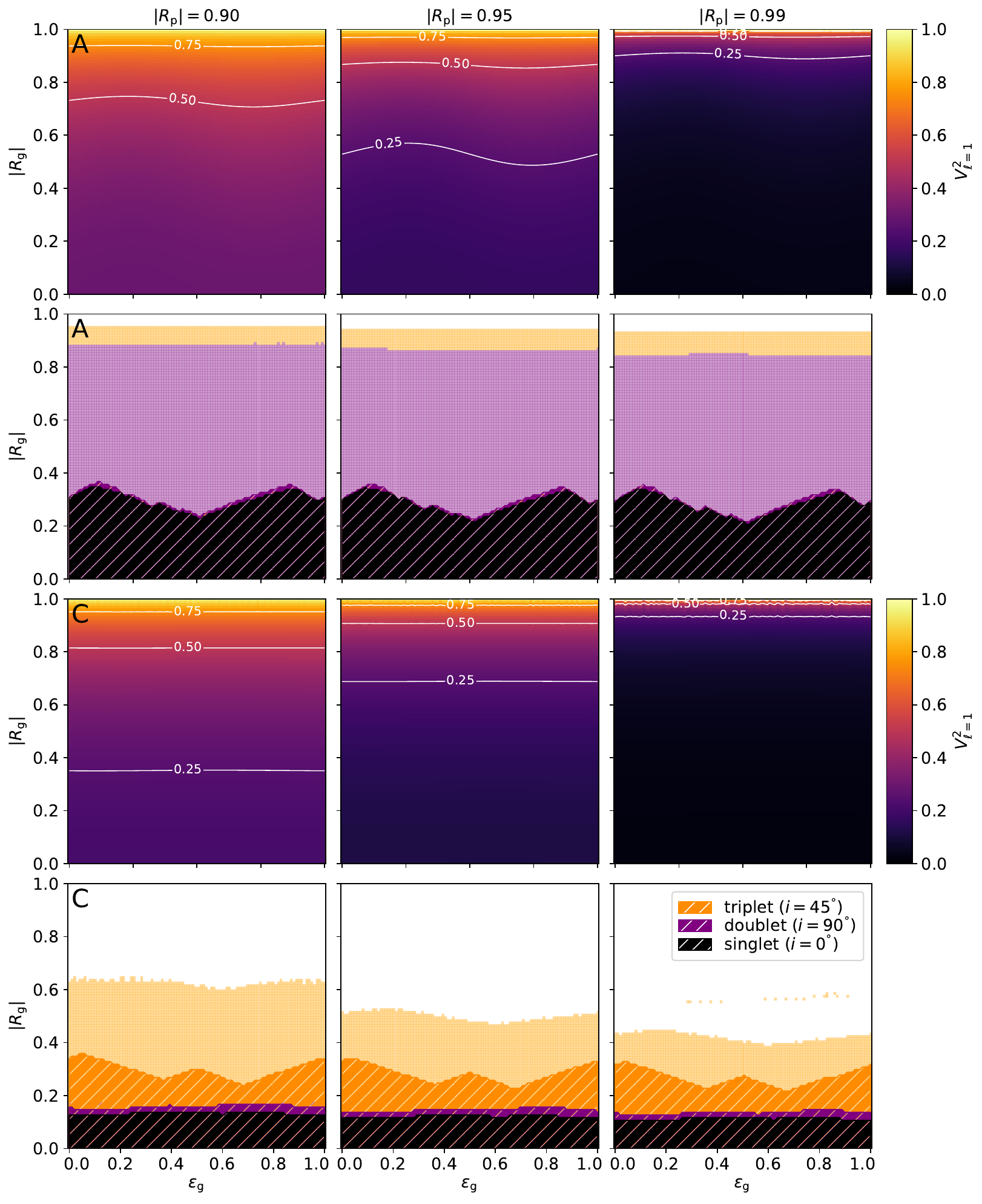}}
    \caption{Same as Fig. \ref{fig: 3x3 epsilon p}, now as a function of the phase shift in the g-mode cavity.}
    \label{fig: 3x3 epsilon g}
\end{figure*}

Here, we show additional figures illustrating the dependence of the visibility and detectability of the mixed mode and multiplet signatures on the coupling strength $q$ (Figs. \ref{fig: 3x3 visibility coupling} and \ref{fig: 3x3 mixed mode detectability coupling}) and on the phase shifts in the p- and g-mode cavity $\epsilon_{\rm p}$ and $\epsilon_{\rm g}$ (Figs. \ref{fig: 3x3 epsilon p} and \ref{fig: 3x3 epsilon g}). The figures presented in this section illustrate that $q$, $\epsilon_{\rm p}$, and $\epsilon_{\rm g}$ can influence both the visibility and detectability of the mixed mode and multiplet signatures. This can further complicate the interpretation of features in the observed PSD with unusually low multipole mode amplitudes.

\section{Additional figures} \label{app: Additional figures}

In this section, we discuss the most important trends trends using the families of colored dots in Figs. \ref{fig: 3x3 visibility} and \ref{fig: 3x3 mixed mode detectability}.
If the reader is interested in the PSD of a particular combination of parameters not shown here, we recommend that they plot it themselves. To do this, it is sufficient to calculate Eq. \eqref{eq: normalized power spectrum 2 cavities} for the relevant parameter combination and multiply the result by a Gaussian, as described in Sect. \ref{sect: relative power spectrum}. The input parameters are given in the figure captions and axes, as well as in Table \ref{tab: sets of stellar parameters}.

\subsection{Additional figure for Sect. \ref{sect: results parameter study}}

\begin{figure}[]
    
    \resizebox{\hsize}{!}{\includegraphics{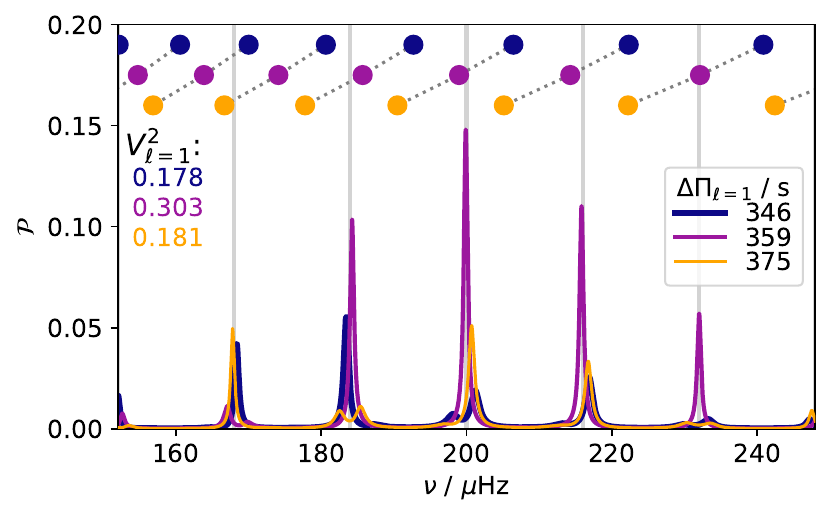}}
    \caption{Relative power spectrum of the dipole modes as a function of frequency for the set of stellar parameters A for different values of the period spacing and $|R_{\rm p}| = 0.95$ and $|R_{\rm g}| = 0.5$. The corresponding visibilities are shown on the left. Gray vertical lines indicate the location of the nominal p-modes. Colored dots indicate the location of the nominal g-modes (their value on the vertical axis has no meaning). The dotted lines connect g-modes of the same radial order. The parameters corresponding to the three power spectra shown here are highlighted as cyan dots in Fig. \ref{fig: 3x3 visibility}.}
    \label{fig: ridges delta Pi}
\end{figure}

Figure \ref{fig: ridges delta Pi} shows that the mixed modes are more pronounced in the PSD when the frequency of the nominal p-modes is approximately equal to that of a nominal g-mode. If this is the case for the majority of the observable pressure radial orders, the visibility is high. If this is not the case, the visibility is low. Since we are fixing all quantities that determine the relative position of the nominal p- and g- modes, with the exception of the period spacing, the ridges appear as a function of $\Delta\Pi_{\ell=1}$. The dependence of $V^2_{\ell=1}$ on $\Delta\Pi_{\ell=1}$ disappears for $|R_{\rm g}| \rightarrow 0$ or 1.

\subsection{Additional figures for Sect. \ref{sect: results detectability of mixed modes}}

\begin{figure}[]
    
    \resizebox{\hsize}{!}{\includegraphics{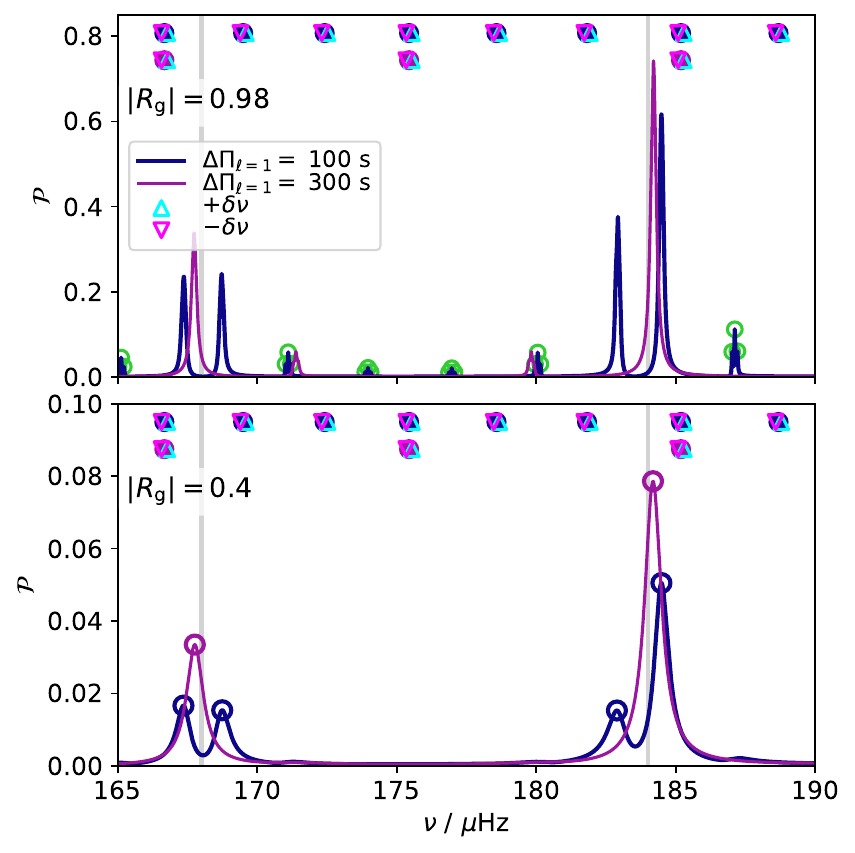}}
    \caption{Relative power spectrum of the dipole modes as a function of frequency for the set of stellar parameters A and two values of the period spacing. The other parameters are $\delta\nu = 0.1\ \mu$Hz, $|R_{\rm p}| = 0.95$, and $i=45^\circ$. The {\it upper panel} corresponds to $|R_{\rm g}| = 0.98$ and the {\it lower panel} to $|R_{\rm g}| = 0.4$. 
    Gray vertical lines indicate the location of the nominal p-modes. Colored dots indicate the location of the nominal g-modes (their value on the vertical axis has no meaning). Triangles indicate the location of the g-modes with $m = \pm 1$. In the upper panel, green circles mark resolved modes that are part of a multiplet (i.e., they do not mark resolved modes that are not part of a multiplet). In the lower panel, there are no multiplets and the colored circles mark all resolved mixed modes. The parameters corresponding to the two power spectra shown here are highlighted as cyan dots in Fig. \ref{fig: 3x3 mixed mode detectability}.}
    \label{fig: resolution delta Pi 2}
\end{figure}
\begin{figure}[]
    
    \resizebox{\hsize}{!}{\includegraphics{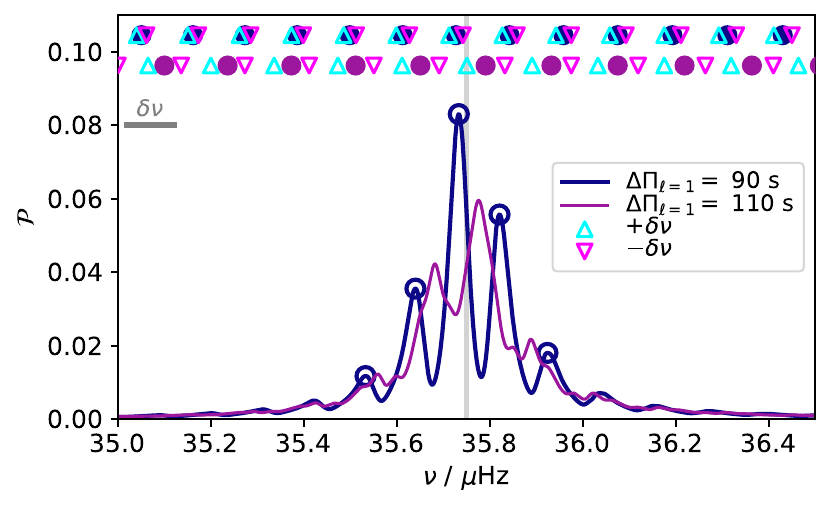}}
    \caption{Relative power spectrum of the dipole modes as a function of frequency for the set of stellar parameters B for two values of the period spacing. The other parameters are $\delta\nu = 0.1\ \mu$Hz, $|R_{\rm p}| = 0.95$, $|R_{\rm g}| = 0.5$, and $i=45^\circ$. The displayed frequency range covers the position of the nominal p-mode at the far left of the PSD. 
    The gray vertical line indicates the location of the nominal p-mode. Colored dots indicate the location of the nominal g-modes (their value on the vertical axis has no meaning). Triangles indicate the location of the g-modes with $m = \pm 1$. Colored circles mark the resolved mixed modes. The parameters corresponding to the two power spectra shown here are highlighted as green dots in Fig. \ref{fig: 3x3 mixed mode detectability}.}
    \label{fig: resolution delta Pi}
\end{figure}
\begin{figure}[]
    
    \resizebox{\hsize}{!}{\includegraphics{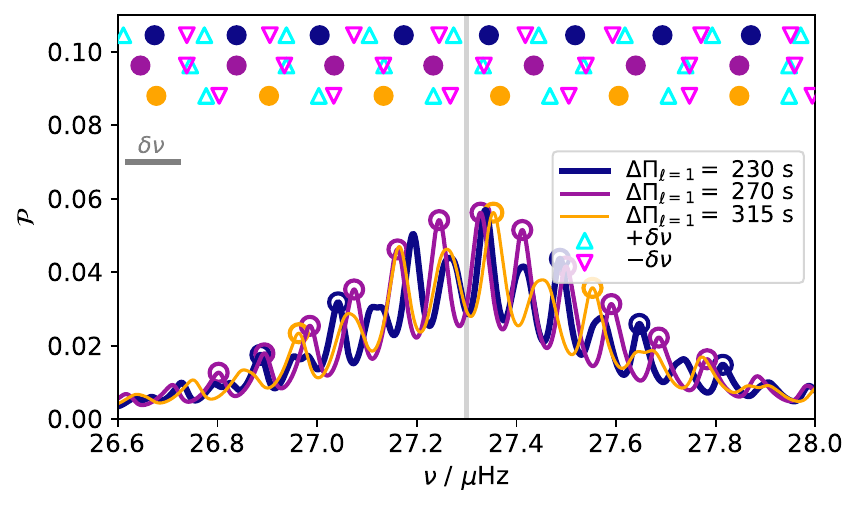}}
    \caption{Relative power spectrum of the dipole modes as a function of frequency for the set of stellar parameters C for two values of the period spacing. The other parameters are $\delta\nu = 0.1\ \mu$Hz, $|R_{\rm p}| = 0.9$, $|R_{\rm g}| = 0.55$, and $i=45^\circ$. The displayed frequency range encompasses the region around a single nominal p-mode. 
    The gray vertical line indicates the location of the nominal p-mode. Colored dots indicate the location of the nominal g-modes (their value on the vertical axis has no meaning). Triangles indicate the location of the g-modes with $m = \pm 1$. Colored circles mark the resolved mixed modes. The parameters corresponding to the two power spectra shown here are highlighted as magenta dots in Fig. \ref{fig: 3x3 mixed mode detectability}.}
    \label{fig: resolution delta Pi 3}
\end{figure}

We first consider the cyan dots in the first row of Fig. \ref{fig: 3x3 mixed mode detectability} to investigate why a smaller value of $\Delta\Pi_{\ell=1}$ seems to favor the presence of both mixed mode and multiplet signatures. Their power spectrum is shown in Fig. \ref{fig: resolution delta Pi 2}. In all figures shown in this section, we focus on the triplets (i.e., $i=45^\circ$).
In the upper panel of Fig. \ref{fig: resolution delta Pi 2}, we see that the multiplet signature is not detectable for most pressure-dominated modes, because it is imprinted in the contribution of the g-mode cavity to the total wave. 
When testing different values of $\Delta\Pi_{\ell=1}$, we find that the multiplet signature of the gravity-dominated modes is only visible if $\Delta\Pi_{\ell=1}$ is small. This can be understood as follows. The most important consequence of a smaller value of $\Delta\Pi_{\ell=1}$ is the larger number of mixed modes in the observable frequency range, which we assume to range from $\nu_{\rm left}$ to $\nu_{\rm right}$. Therefore, the gravity-dominated mixed modes are located at frequencies that are much closer to those of their corresponding nominal g-modes. The smaller difference between the frequencies of the mixed modes and the nominal g-modes favors the detectability of the multiplet signature, since the mixed modes can retain more of their g-mode character.
In the lower panel of Fig. \ref{fig: resolution delta Pi 2}, we see that at smaller values of $|R_{\rm g}|$, only the most pressure-dominated mixed modes are recognizable. Since there are more pressure-dominated modes at smaller $\Delta\Pi_{\ell=1}$ due to the larger number of mixed modes, more than one mixed mode per pressure radial order is recognizable at smaller $\Delta\Pi_{\ell=1}$ and the mixed mode signature is detectable.

Next, we consider the green dots in the second row of Fig. \ref{fig: 3x3 mixed mode detectability} to investigate the two abrupt changes between the two dots. Their power spectra are shown in Fig. \ref{fig: resolution delta Pi}.
The first change is that the multiplet signature can no longer be distinguished from the mixed mode signature for the lower value of $\Delta\Pi_{\ell=1}$. This can be clearly seen in Fig. \ref{fig: resolution delta Pi}, as the frequencies of the nominal g-modes shifted by $\delta\nu$ ($m=\pm1$) almost overlap with those of the unshifted g-modes ($m=0$) for the lower value of $\Delta\Pi_{\ell=1}$.
The second change is that the mixed mode signature suddenly becomes recognizable again for the lower value of $\Delta\Pi_{\ell=1}$. This is also because frequencies of the modes shifted by $\delta\nu$ now almost coincide with those of the unshifted g-modes, thus reducing the number of observable peaks.
In other words, the frequencies of the mixed mode multiplets align with those of other multiplets, thus simplifying the observed oscillation pattern and making the mixed modes easier to detect.
For the higher value of $\Delta\Pi_{\ell=1}$, there are so many pressure-dominated mixed modes due to the triplets that the mixed mode signature is blurred. For the lower value, there are fewer peaks with greater height in the PSD, thus making them more resistant to clumping.

In Fig. \ref{fig: resolution delta Pi 3}, we examine the PSD corresponding to the magenta dots in the bottom row of Fig. \ref{fig: 3x3 mixed mode detectability}. In particular, we want to answer the question of why the detectability of the multiplet signature is less favorable when $\Delta\Pi_{\ell=1}$ is close to 230 s or 315 s.
Following a similar logic to that in Fig. \ref{fig: resolution delta Pi}, Fig. \ref{fig: resolution delta Pi 3} shows that mode multiplets are easier to detect when their frequencies are close to those of other modes, as this reduces the apparent number of peaks in the PSD. When the frequencies of the peaks are slightly offset, as is the case for $\Delta\Pi_{\ell=1} = 230$ s and 315 s, the peaks are broader and less pronounced, thus making them less recognizable. This conceals the multiplet signature.

\end{appendix}

\end{document}